\documentclass[twocolumn]{aastex62}
\pdfoutput=1
\usepackage{amsmath,amstext,multirow}
\usepackage[T1]{fontenc}
\usepackage{apjfonts,xcolor}
\usepackage[figure,figure*]{hypcap}
\usepackage{enumitem}

\def\be{\begin{eqnarray}}   \def\ee{\end{eqnarray}}
\def\ben{\begin{equation}\begin{aligned}} \def\een{\end{aligned}\end{equation}}
\shorttitle{K2GAP target selection}
\shortauthors{Sharma et al.}
\newcommand{\kepler}{\textit{Kepler}}

\newcommand{\numax}{$\nu_{\rm{max}}$}
\newcommand{\dnu}{$\Delta\nu$}

\begin{document}

\title{The K2 Galactic Archaeology Program: Overview, target selection and survey properties}
\author{Sanjib Sharma}
\affiliation{Sydney Institute for Astronomy, School of Physics, The University of Sydney, NSW 2006, Australia}
\affiliation{ARC Centre of Excellence for All Sky Astrophysics in Three Dimensions (ASTRO-3D)}
\author{Dennis Stello}
\affiliation{School of Physics, University of New South Wales, Sydney, NSW 2052, Australia}
\affiliation{Sydney Institute for Astronomy, School of Physics, The University of Sydney, NSW 2006, Australia}
\affiliation{Stellar Astrophysics Centre, Department of Physics and Astronomy, Aarhus University, DK-8000 Aarhus C, Denmark}
\affiliation{ARC Centre of Excellence for All Sky Astrophysics in Three Dimensions (ASTRO-3D)}
\author{Joel C. Zinn}
\affiliation{School of Physics, University of New South Wales, Sydney, NSW 2052, Australia}
\author{Joss Bland-Hawthorn}
\affiliation{Sydney Institute for Astronomy, School of Physics, The University of Sydney, NSW 2006, Australia}
\affiliation{ARC Centre of Excellence for All Sky Astrophysics in Three Dimensions (ASTRO-3D)}

\begin{abstract}
K2 was a community-driven NASA mission where all targets were proposed through guest observer programs. Among those, the K2
Galactic Archaeology Program (K2GAP) was devoted to measuring asteroseismic signals from giant stars to inform studies of the Galaxy, with about 25\% of the observed K2 targets being allocated to this program.  Here, we provide an overview of
this program.
We discuss in detail the target selection procedure
and provide a python code that implements the selection function (\url{github.com/sanjibs/k2gap}).
Additionally, we discuss the detection completeness of the asteroseismic parameters $\nu_{\rm max}$ and $\Delta \nu$.
Broadly speaking, the targets were selected based on 2MASS color $J-Ks > 0.5$,
with finely tuned adjustments for each campaign.
Making use of the selection function we compare the observed distribution of asteroseismic masses to theoretical predictions. The median asteroseismic mass is higher by about 4\%  compared to predictions.
Additionally, the number of seismic detections is on average 14\% lower than expected.
We provide a selection-function-matched mock catalog of stars based on a synthetic model of the Galaxy for the community to be used in subsequent analyses of the
K2GAP data set (\url{physics.usyd.edu.au/k2gap}).
\end{abstract}


\section{Introduction}
The use of asteroseismology to inform studies of the Milky Way were proposed over a decade ago \citep{2009A&A...503L..21M},
when it became possible  for the first time  to detect oscillations in hundreds, even thousands, of distant stars using
continuous high-cadence, photometric data from space-based missions; particularly from CoRoT and \kepler\ \citep[e.g.][]{2009Natur.459..398D,2013ApJ...765L..41S}.
However, these early missions were far from ideal for studying the Galaxy as a whole, and indeed were never designed with this in mind \citep{2016ApJ...822...15S, 2017ApJ...835..163S}.
The main reasons were the limited sky coverage and the lack of a well-defined target selection function. A well understood selection function is fundamental if we are to make meaningful comparisons between observed and synthetic/modeled stellar populations \citep{2011ApJ...730....3S}. Such studies play an important role in making robust inferences about the Galactic stellar populations.

\begin{figure*}[tb]
\centering \includegraphics[width=0.99\textwidth]{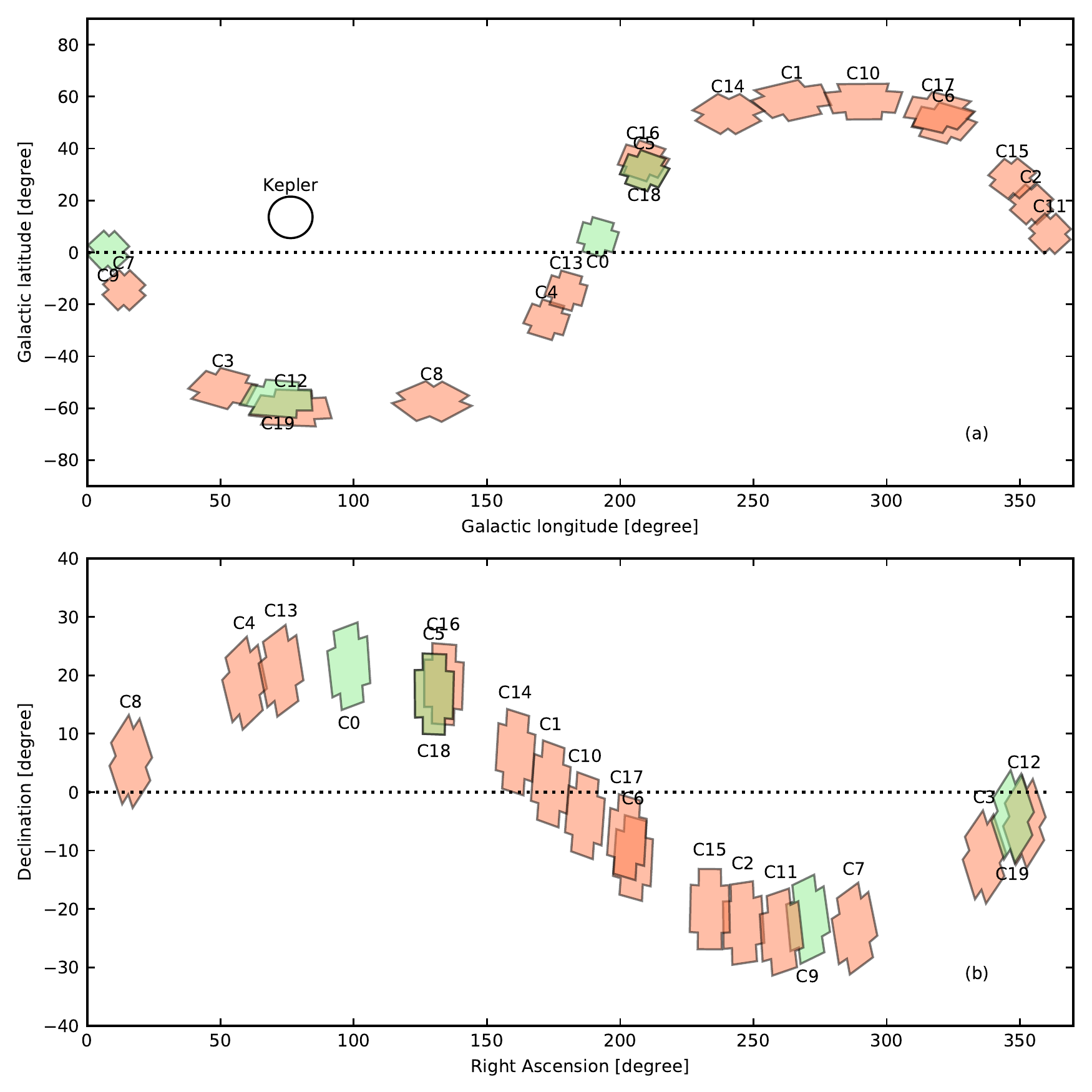}\caption{Footprint of the different K2 campaigns in Galactic and equatorial coordinates.  Campaigns C0, C9, C18, and C19, are not suitable for selection-function-based studies and are shown in green. The first three do not follow a well defined color-magnitude selection while C19 has very few seismic detections.
\label{fig:fov_all_campaigns}}
\end{figure*}

A decade ago, the primary NASA {\it Kepler} mission targetted a single field in the Cygnus and Lyra constellations for four years. In 2014, the compromised satellite was then re-purposed for a new mission to stare for $\sim 80\,$ days at a time in different directions along the ecliptic (\autoref{fig:fov_all_campaigns});  the so-called ``K2 mission''.
K2 was now able to probe stellar populations in many directions, covering much more of the Galaxy than achieved in earlier missions. This included the halo, the bulge, the thin and thick disks, and at vastly different Galactic radii and heights above and below the plane.
To take advantage of this opportunity, the K2 Galactic Archaeology Program (K2GAP) was formed around an international collaboration with the aim of detecting oscillations in thousands of red giants along the ecliptic.

The primary goal was to establish robust stellar ages for the major Galactic stellar components \citep{2019MNRAS.490.4465R,2019MNRAS.490.5335S}, and to significantly improve on what was possible with ESA's {\it Gaia} and ground-based spectroscopic survey data alone \citep{2019MNRAS.486.1167B}.
This was achieved by devising a well-documented, reproducible target selection to eliminate the limitations of previous space-based seismic observations, and to maximize the synergy with Galactic spectroscopic surveys.
In total, K2 provided 18 full-length observing campaigns before the fuel ran out and the spacecraft was retired by the end of 2018.
The K2GAP survey has already published a ``proof of concept'' study \citep{2015ApJ...809L...3S} along with a succession of well-tested data products.
These are Data Release 1 (DR1) containing seismic results from campaign 1 (C1) \citep{2017ApJ...835...83S}, Data Release 2 (DR2) containing seismic results for C3, C6, and C7 \citep{2020ApJS..251...23Z}, and the final Data Release 3 (DR3) with results from all campaigns
in one homogeneous catalog \citep{2021arXiv210805455Z}. In addition, science results were published by \citet{2019MNRAS.490.4465R}, \citet{2019MNRAS.490.5335S,2020arXiv200406556S} and \citet{2021arXiv210805455Z}.

The overarching vision for the K2GAP seismic data was always
that it should be combined with complementary data from the
many other stellar surveys that emerged in the same decade.
In recognition of K2GAP's potential, several large ground-based, spectroscopic surveys have targetted K2GAP fields.  The surveys with the largest K2GAP overlap are APOGEE, K2-HERMES, and LAMOST.
As we show, this synergy has proved to be very effective in determining
improved stellar ages for thousands of stars, an important goal for Galactic archaeology \citep{2002ARA&A..40..487F}.

In this paper we report the target selection for each K2 campaign in detail. We discuss detection completeness of the asteroseismic parameters $\nu_{\rm max}$ and $\Delta \nu$ for the oscillating giants.
We make a detailed comparison of the observed distribution of stars in
$\nu_{\rm max}$,
$V_{JK}$, and $\kappa_{M}$ with that of selection-matched mock catalogs. Finally, we discuss the implications of our results for asteroseismic relations and Galactic archaeology.

\begin{figure}[tb]
\centering \includegraphics[width=0.49\textwidth]{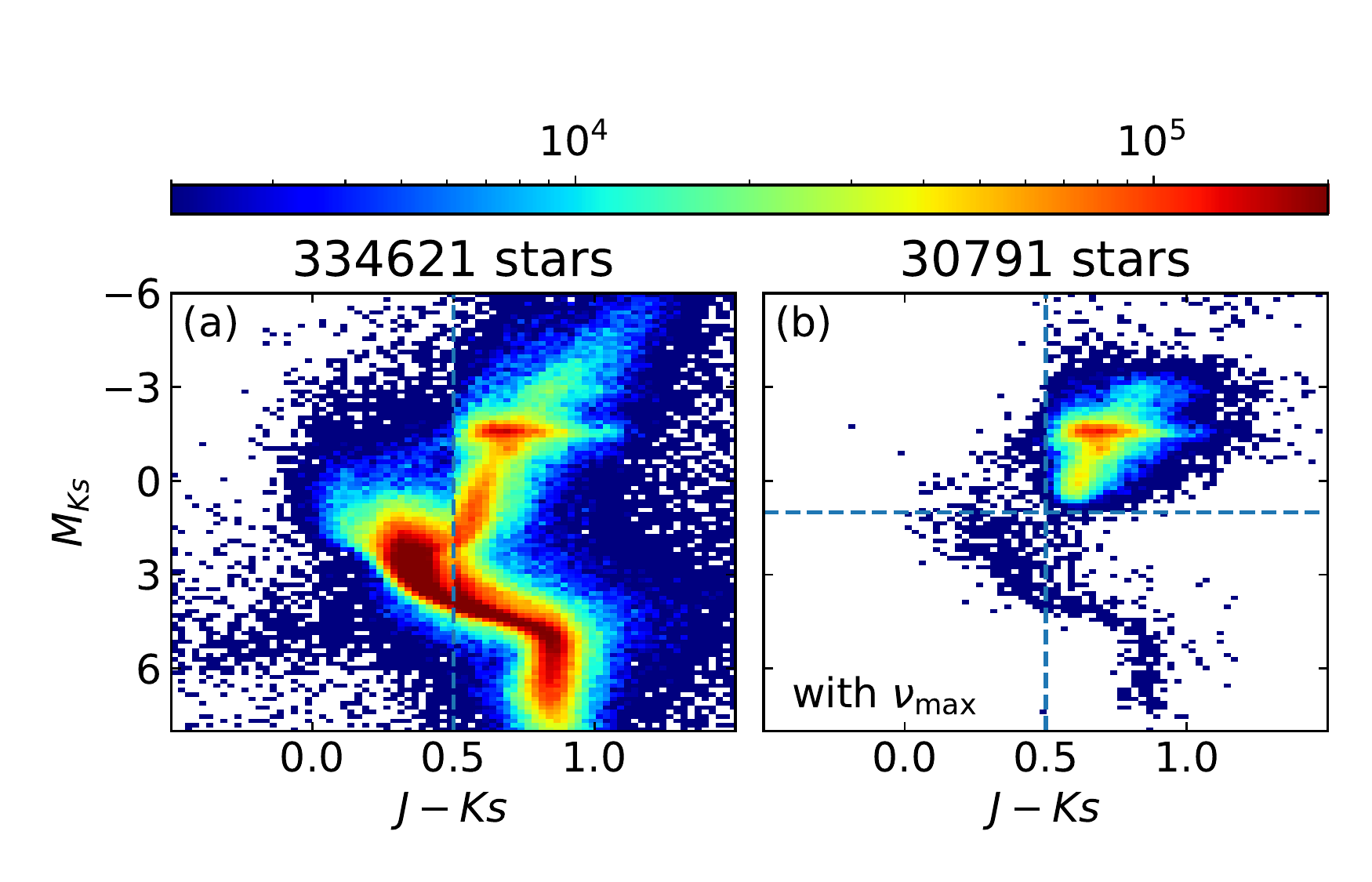}\caption{(a) Color and absolute magnitude distribution of all stars observed by K2 in 30 minute cadence that have a Gaia EDR3 parallax.
The absolute magnitude is estimated using Gaia EDR3 parallax.
The vertical line $J-Ks=0.5$ marks the K2GAP color selection.
The number of stars in each bin is indicated by the color bar. (b) The subset of stars with asteroseismic detection of \numax. Most of the stars with \numax\  detections have $J-Ks>0.5$. Stars with \numax\ detected and $M_{Ks}>1$ have either incorrect $\nu_{\rm max}$ or parallax.
\label{fig:color_absmag}}
\end{figure}

\begin{figure}[tb]
\centering \includegraphics[width=0.49\textwidth]{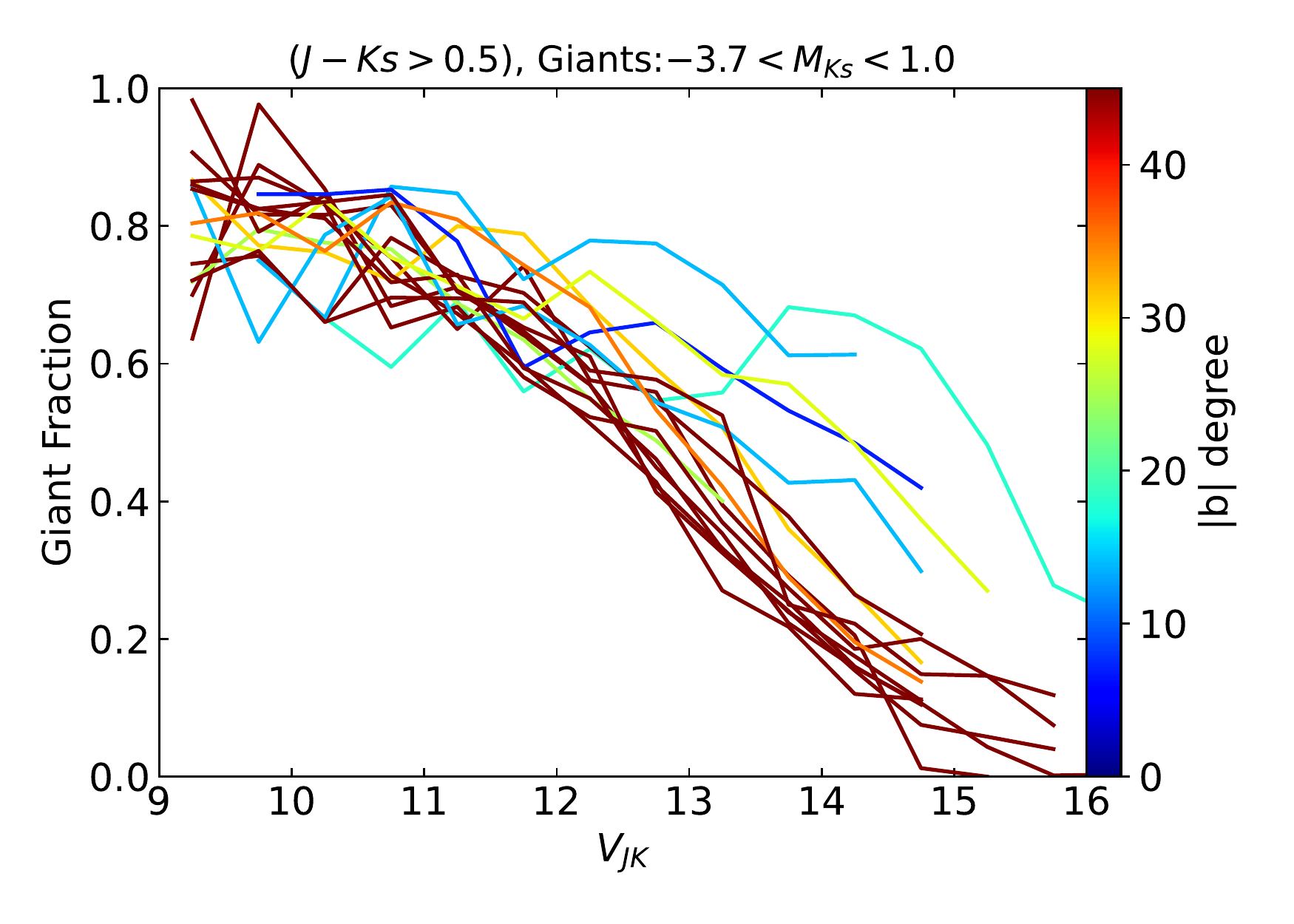}\caption{The fraction of stars proposed by K2GAP that are giants as a function of apparent magnitude $V_{JK}$ for different K2 campaigns.
The stars satisfy the selection function given in \autoref{tab:selfunc}.
Giants were identified based on their absolute magnitude being  $-3.7<M_{Ks}<1$. $M_{Ks}$ was estimated using
the {\it Gaia} parallaxes.
\label{fig:giant_fraction}}
\end{figure}

\section{Method}
\subsection{Primary target selection strategy}
Our target selection strategy was designed to be easily
reproducible, which aids the study of ensembles rather than individual stars.
This is especially important for Galactic archaeology where we need to fit Galactic models to the observational data
by taking the target selection into account \citep{2014ApJ...793...51S}, but is also useful for exoplanet population studies.

A necessary first step for selecting targets is to have
an input catalog that is well understood, has reliable
photometry and covers all regions of the sky that we are interested in.
With this in mind, we adopt the 2MASS all-sky catalog because of its robust photometry and completeness in the magnitude range we are targeting ($9<V_{JK}<16$).
A color limit of $(J-Ks)>0.5$ was adopted throughout the survey. This was designed to focus on red giants; our primary asteroseismic targets. It can be seen in \autoref{fig:color_absmag}
that most of the stars for which we can detect oscillations have $(J-Ks)>0.5$ and $M_{Ks}<1$. Stars with $(J-Ks)<0.5$ are typically dwarfs ($M_{Ks}>1$) that have oscillation frequencies that are too large to be detected by the 30-minute cadence of K2. A color limit of $(J-Ks)>0.5$ excludes some giants,
such as the blue extension of the red clump stars (the so-called horizontal branch),
which is predominantly metal-poor and are rare.

Although reduced proper motions have been used to separate dwarfs from giants in the past, we avoided it for multiple reasons.  First, it introduces a kinematic bias, which is undesirable for Galactic archaeology.  Secondly, the pre-$Gaia$ proper motions available (UCAC) at the time of the K2GAP selection, had significant uncertainties.  Finally, the dwarfs that we were not interested in were desirable for exoplanet studies; and a simple selection function would also benefit those.  In summary, the simplest color based selection criteria was found to be the best suited for both Galactic archaeology and exoplanet population studies.

In addition to color, stars were restricted in apparent magnitude (brightness) between a lower and upper limit. The lower (bright) limit was
chosen to avoid overly saturated stars, while the upper (faint) limit was
chosen to avoid observing ensembles with a low yield of oscillating giants.
For the first 3 campaigns (C1, C2, and C3) we used the 2MASS $H$ magnitude to select in brightness. However, for the later campaigns, we adopted
\begin{equation}
V_{JK}=Ks + 2.0 (J-Ks+0.14)+0.382\exp[(J-Ks-0.2)/0.50],
\label{equ:vjk}
\end{equation}
which is an approximation of $V$ band magnitude measured from 2MASS $J$ and $K$ bands. As shown in \citet{2018MNRAS.473.2004S}, the formula is accurate to 0.05 dex for stars in
$4500<T_{\rm eff}/{\rm K}<6500$.
We adopted $V_{JK}$ for the later campaigns because K2 collects data in $Kp$ band, which is significantly bluer than $H$ band. Additionally, spectroscopic surveys like K2-HERMES \citep{2018MNRAS.473.2004S} and LAMOST that were following up the K2 targets, observe in the $V$ band.

A bright magnitude limit of $V_{JK}>9$ ($H>7$ for C1, C2, and C3) was adopted.
However, the faint limit was different for each campaign, with the typical limit being $V_{JK}=15$. The limit for each campaign was determined by weighing the potential scientific return versus the drop in yield of oscillating giants when going towards fainter magnitudes.
The drop in yield of oscillating giants when going fainter
is due to a couple of reasons. First, in the Galaxy the overall fraction of stars that are giants for a given apparent magnitude drops as we go to fainter magnitudes. This is shown in \autoref{fig:giant_fraction}. Secondly, fainter stars have
lower signal-to-noise ratio, which hinders our ability to detect the oscillations. Based on
simulations the incompleteness for the K2 observations was predicted to set in at around $V_{JK}=14$; see \autoref{sec:numax_vmag_dist} for more details. This prompted us to not propose targets in general that were too faint.

\begin{table*}[tb!]
\caption{ The selection function of K2GAP targets for each campaign. The first column is campaign number while the color-magnitude selection is given in the last column (eighth).
Columns two to seven list the number of stars for a given data set, with the set to the right being a subset of the set to its left.
The second column lists stars observed by K2 (\texttt{epic\_id > 201000000}). The third column lists stars observed by K2 that are in the K2GAP target list (\texttt{flag\_ga = 1}). The fourth column lists stars that are K2GAP complete, meaning it excludes serendipitous observations of K2GAP targets from other programs.
The fifth column lists stars that satisfy a strict  color-magnitude selection function, which is given in the eighth column. The sixth column lists the completeness fraction of stars in the fifth column.
The seventh column lists stars for which we have a $\nu_{\rm max}$ measurement in the range 10 to 270 $\mu$Hz. Finally, the eighth column is the selection function. All stars effectively have $J<15$ due to 2MASS quality cuts.}
\begin{tabular}{@{} l  r r r r l r l}
\hline
C & K2 & GAP & GAP & GAP & $f_{\rm comp}$ & $\nu_{\rm max}$ & Selection function (SF)\\
& Obs & Obs  & Com &  SF & SF & SF & $(J<15)\&$  \\
\hline
\hline
0 &   7748 &    452 &      0 &      0 &  0.000 &      0 &                                \\
1 &  21646 &   8630 &   8409 &   8398 &  0.983 &   1151 & $((J-Ks) \geq 0.5)\&(7 \leq H < 12.927)$ \\
2 &  13401 &   5138 &   3924 &   3465 &  0.904 &   1140 & $((J-Ks) \geq 0.5)\&(7 \leq H < 11.5)\&(c \in \{17,12,6,14,10\})$ \\
3 &  16375 &   3904 &   3450 &   3407 &  0.800 &   1069 & $((J-Ks) \geq 0.5)\&(7 \leq H)\&[(H < 12.0)\&(c \geq 0)]$ \\
4 &  15853 &   6357 &   4938 &   4937 &  0.993 &   1931 & $((J-Ks) \geq 0.5)\&(9 \leq V_{JK} < 13.447)$ \\
5 &  25137 &   9829 &   9829 &   9820 &  0.991 &   2666 & $((J-Ks) \geq 0.5)\&(9 \leq V_{JK} < 15.0)$ \\
6 &  28289 &   8313 &   8311 &   8303 &  0.995 &   2193 & $((J-Ks) \geq 0.5)\&(9 \leq V_{JK} < 15.0)$ \\
7 &  13483 &   4362 &   4085 &   4085 &  0.996 &   1678 & $((J-Ks) \geq 0.5)\&[(14.276 \leq V_{JK} < 14.5)\&(c = 14)\ ||\ $ \\
&        &        &        &        &        &       & $(9 \leq V_{JK} < 14.5)\&(c \in \{6,17\})]$ \\
8 &  24187 &   6186 &   5383 &   4392 &  0.995 &    954 & $((J-Ks) \geq 0.5)\&[ (9 \leq V_{JK} < 14.5)\ ||\  (J-Ks)<0.7)\&(14.5 < V_{JK} < 14.58)]$ \\
9 &   1751 &      0 &      0 &      0 &  0.000 &      0 &                                \\
10 &  28345 &   8947 &   8559 &   7382 &  0.995 &   1196 & $((J-Ks) \geq 0.5)\&[(9 \leq V_{JK} < 14.5)\ ||\  (J-Ks)<0.7)\&(14.5 < V_{JK} < 15.577)]$ \\
11 &  14209 &   4344 &   3403 &   2701 &  0.986 &    449 & $((J-Ks) \geq 0.5)\&(9 \leq V_{JK})\&[(V_{JK} < 15.0)\&(c=3)\ ||\ (V_{JK} < 14.5)\&(c=2)\ ||\ $ \\
&        &        &        &        &        &       & $( V_{JK} < 14.175)\&(c=8)]$   \\
12 &  29221 &  14013 &  14013 &  13019 &  0.967 &   1167 & $((J-Ks) \geq 0.5)\&(9 \leq V_{JK} < 16.0)$ \\
13 &  21434 &   5973 &   4686 &   4381 &  0.982 &   1597 & $((J-Ks) \geq 0.5)\&(9 \leq V_{JK})\&[(V_{JK} < 15.0)\&(c=3)\ ||\ (V_{JK} < 14.5)\&(c=8)\ ||\ $ \\
&        &        &        &        &        &       & $( V_{JK} < 14.0)\&(c \in \{0,4,7,9,12,13\})\ ||\ (V_{JK} < 12.82)\&(c=14)]$ \\
14 &  29897 &   7134 &   5965 &   5587 &  0.980 &   1063 & $((J-Ks) \geq 0.5)\&(9 \leq V_{JK} < 15.0)\ ||$ \\
15 &  23278 &   7625 &   7000 &   5820 &  0.987 &   2254 & $((J-Ks) \geq 0.5)\&(9 \leq V_{JK})\&[(V_{JK} < 15.5)\&(c=3)\ ||\ ( V_{JK} < 15.0)\&(c=8)$ \\
&        &        &        &        &        &       & $(V_{JK} < 14.5)\&(c \in \{2,4,7,9,12,13,14\})\ ||\ (V_{JK} < 13.838)\&(c=5)]$ \\
16 &  29888 &  10672 &   9581 &   7798 &  0.976 &   1805 & $((J-Ks) \geq 0.5)\&(9 \leq V_{JK} < 15.0)\ ||$ \\
17 &  34398 &   7124 &   3003 &   2042 &  0.980 &    672 & $((J-Ks) \geq 0.5)\&(9 \leq V_{JK})\&[ (V_{JK} < 16.0)\&(c \in \{3,8\})\ ||\ $ \\
&        &        &        &        &        &       & $\ ( V_{JK} < 12.414)\&(c \notin \{3,8\})]$ \\
18 &  20427 &   3164 &      4 &      0 &  0.000 &      0 &                                \\
19 &  33863 &  10030 &   6248 &   5882 &  0.977 &      0 & $((J-Ks) \geq 0.5)\&(9 \leq V_{JK} < 14.8)$ \\
\hline
All & 432830 & 132197 & 110791 & 101419 &  0.974 &  22985 & \\
\end{tabular}
\label{tab:selfunc}
\tablecomments{The circular pointing identifier $c$ is shown in \autoref{fig:fovC1}. For campaign numbers greater than or equal to 10, the CCD module corresponding to $c=1$ is broken.}
\end{table*}

\begin{figure}[tb]
\centering \includegraphics[width=0.49\textwidth]{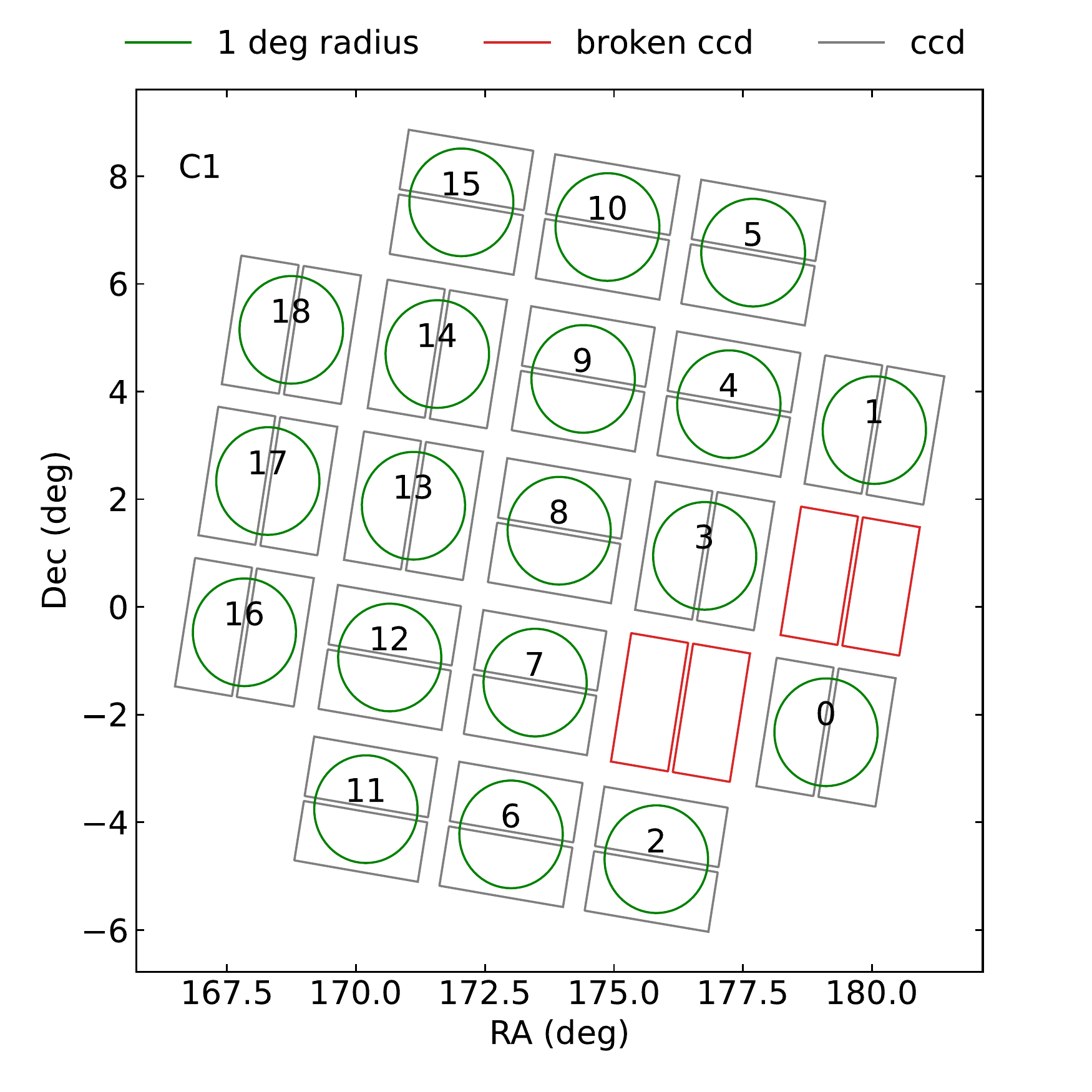}\caption{
Field of view of a typical K2 campaign. Shown here is campaign C1. The solid lines outline the 42 CCDs arranged in 21 modules. The broken modules are shown in red. The green circle shows the 1 degree radius field of view of the HERMES spectrograph. Note, the module
numbers defined by us is similar but not the same as the official module numbers; our numbering starts from 0 and we exclude the four corner modules(1, 5, 21 and 25) and the two broken modules (3 and 7).
\label{fig:fovC1}}
\end{figure}

\begin{figure*}[tb]
\centering \includegraphics[width=0.99\textwidth]{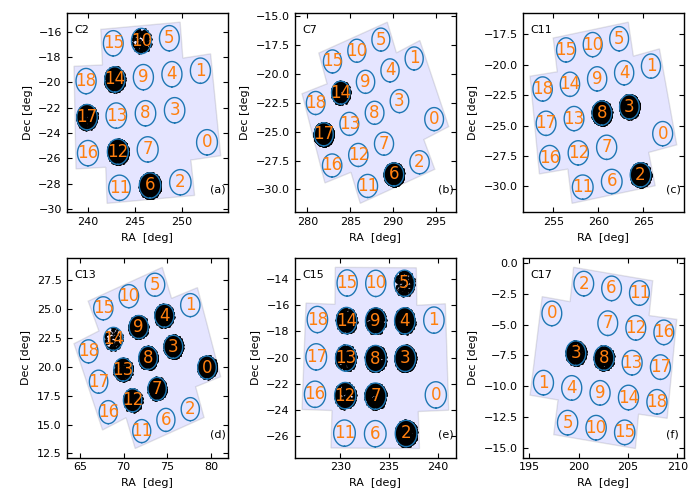}\caption{Distribution of K2GAP stars observed by K2 in campaigns where stars were selected to lie in one degree radius circles. Each campaign has 19 one degree circles, numbered 0 to 18, located at the centers of each CCD module (excluding the two broken CCD modules).
\label{fig:fov_cricno}}
\end{figure*}

\subsection{Field of view} \label{sec:fov}
For certain studies it is important to know  the completeness of the observed sample as well as the foot print of the field of view.
The field of view of K2 is 115.64 square degrees comprising a mosaic of 21 CCD modules each made up of two 1024x2200 pixel CCDs (3.98 arcsec pixel scale) with slight gaps between the two CCDs and between the 21 modules (see  \autoref{fig:fovC1}).
Although the actual area of a module is 5.507 square degrees, the proposed stars were confined to 5.482 square degrees, most likely due to a CCD-edge buffer of a few pixels introduced by the python package \texttt{K2Fov} \citep{2016ascl.soft01009M}\footnote{\texttt{K2Fov} is a python module provided by the K2 mission team to identify if a target is within the field of view (\url{https://github.com/KeplerGO/K2fov})}, which we used to select the targets.
For certain campaigns with high target density, we proposed stars in 1 degree circles located at the center of selected CCD modules, to make the spectroscopic followup easier (see \autoref{fig:fov_cricno}).
Such a circle has a photosensitive area of 2.961 square degrees (94.251\% of the circle's area), which for this given size is a maximum possible photosensitive area among all circle placements.

\begin{table}[htb!]
\caption{2MASS quality selection criteria}
\begin{tabular}{@{}llll}
\hline
fFlag  & K2GAP & K2-HERMES & Description \\
& criterion & criterion & \\
\hline
\hline
Qflag & $\leq$ 'BBB' & $=$ 'AAA'    & J,H,K photometric quality\\
Bflag & $=$ '111'    & $=$ '111'    & blend flag \\
Cflag & $=$ '000'    & $=$ '000'    & contamination flag\\
Xflag & $=$ 0        & $=$ 0        & Ext source\\
Aflag & $=$ 0        & $=$ 0        & solar system object\\
prox  & $>$ 6 arcsec & $>$ 6 arcsec & distance to nearest star\\
\hline
\end{tabular}
\label{tab:qualsel}
\end{table}

\begin{table*}[htb!]
\caption{Priority order of selecting targets}
\begin{tabular}{@{}llll}
\hline
Order  & Description& Criterion & Used in campaigns\\
\hline
\hline
0 & APOGEE & $((J-Ks) \geq 0.5)\&(\log g < 3.8)$ & [1-19]\\
1 & RAVE & $((J-Ks) \geq 0.5)\&(\log g < 3.8)$ & [1-19] \\
2 & GAIA-ESO & $((J-Ks) \geq 0.5)\&(1.8< \log g < 3.8)$ & [2, 12]\\
3 & MISC & miscellaneous special targets. & [3, 6, 7, 11, 17]\\
4 & SEGUE & $(0.5<(g-r)<1.3)\&(0<\log g<3.5)\&(14<r<19)\&({\rm snr}>15)\&$({\rm flag='nnnnn'}) & [8, 10, 12, 14, 16, 17, 18, 19]\\
5 & K2 $\nu_{\rm max}$ & Stars having $\nu_{\rm max}$ measurements from previous K2 campaigns. & [17, 18, 19] \\
10 & 2MASS primary& $(J-Ks) \geq 0.5$ & [1-19]\\
11 & SDSS $l$-color & $(0.5 \leq (g-r)<0.7)\&(16<r \leq 17)\&(l_{\rm color}>0.1)$ & [8, 10, 12]\\
12 & SDSS $\mu_{\rm reduced}$ & $(0.5 \leq (g-r)<0.7)\&(16<r \leq 17)\&(\mu_{\rm reduced}<5)$ & [8, 10, 12] \\
13 & 2MASS $\mu_{\rm reduced}$ & $((J-Ks) \geq 0.5)\&(14.5 \leq V_{JK} < 16.0)\&(\mu_{\rm reduced}<5)$ & [12, 14, 16]\\
14 & 2MASS secondary& $(0.5 \leq (J-Ks) < 0.7)\&(14.5 \leq V_{JK} < 16.5)$ & [8, 10] \\
10+k & 2MASS in circles & $c=c_{\rm list}[k]$ & [2, 7, 11, 13, 15, 17] \\
\hline
\end{tabular}
\label{tab:priority}
\end{table*}

\subsection{Catalog of targets}
We provide a master catalog of all targets observed by K2 in
the file
\texttt{k2\_observed\_targets.fits} \footnote{\url{http://www.physics.usyd.edu.au/k2gap/download/k2\_observed\_targets.fits}}.
To construct the catalog we start with a list of \texttt{epic\_id}s of observed targets from the {\it Kepler} science website \footnote{\url{
http://keplergo.github.io/KeplerScienceWebsite/}}. It also contains campaign names  \texttt{campaign}, which we convert to an integer \texttt{cno}.
This list includes all targets selected by any of the K2 programs, not just those proposed by the K2GAP.
We first cross matched it to the epic catalog based on \texttt{epic\_id} and added columns from it. To supplement the photometric information from 2MASS, we did our
own cross match with 2MASS instead of relying on the information
provided by the EPIC catalog \citep{2016ApJS..224....2H}.  This was
because the EPIC catalog was missing entries for about 13,000 targets, in spite of targets being stars ($\texttt{epic\_id} > 201000000$). Moreover, the 2MASS columns \texttt{mflg} and \texttt{prox} were incorrect for the majority of the stars in the EPIC. These flags, among others, were used in K2GAP to select stars with good quality photometry (\autoref{tab:qualsel}). The catalog contains
588,991 targets. For the rest of the paper we exclusively focus on the 432,830 targets that have \texttt{epic\_id > 201000000}; the rest being non-stellar targets like asteroids, planets, moons and so on.

To the base catalog we have added two additional columns that provide information about the K2GAP (\texttt{flag\_ga} and \texttt{flag\_sf}).
Targets that are on the K2GAP target list have
\texttt{flag\_ga=1} of which there are 132,197 in total.
This amounts to 30.5\% of all the observed K2 targets. These stars also include the `serendipitous' ones selected by other programs that are in the K2GAP target list. We remove the serendipitous targets to create a K2GAP-chosen sample (referred to as K2GAP complete),  following a procedure that is described in  \autoref{sec:proposed_targets}.
A majority of the  K2GAP complete targets
follow a strict color magnitude selection function (101419), and these can be identified with \texttt{flag\_sf=1}.

\subsection{Peculiarities of certain campaigns}
Due to a late change in roll angle for C3, the actual field of view was slightly shifted compared to the one provided by the \texttt{K2Fov} software before observations. This meant that some of the proposed targets were unobservable.
Later, the permanent failure of one of the CCD modules (marked as  1 in \autoref{fig:fov_cricno}) occurring during C10 observations (after C11, C12, and C13 target selection had been locked in) resulted in selected stars falling on that module being only partly or not at all observed for the affected campaigns.

Campaigns C0, C9, C18, and C19 were all unique and they had very little seismic detections.
C0 was conducted as a full-length engineering test to prove the viability of the K2 mission
and to fine tune the observational setup, such as pointing and choice of aperture size.
C9 was dedicated for gravitational microlensing studies, hence no community targets sought.
C18 and C19 had light curves of shorter duration.
C18 observations were terminated after 50 days due to low fuel. About 3000 K2GAP
targets were observed but all of them were found to be serendipitous selections, hence,
their selection function is not known.
C19 observations only lasted 30 days before the spacecraft ran out of fuel. This campaign also suffered from erratic pointing.

\begin{figure}[tb]
\centering \includegraphics[width=0.49\textwidth]{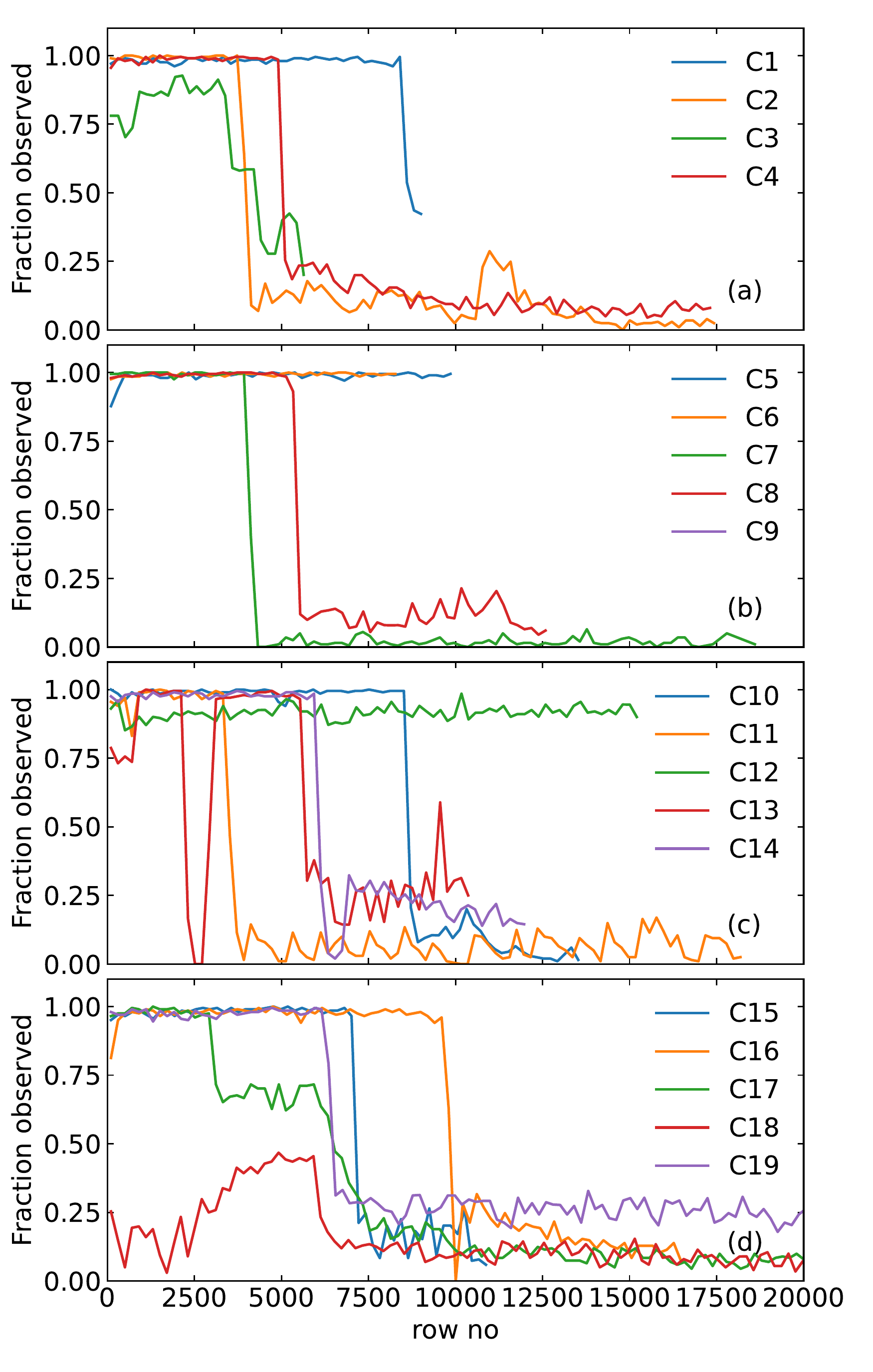}\caption{Fraction of stars observed by K2 as a function of row no in the list of targets proposed by K2GAP. The sharp fall in fraction marks the row number up to which the sample is complete.  For most campaigns, the completeness is typically high, greater than 0.97. C3 and C12 have slightly lower completeness, 0.85 and 0.91 respectively.  For C3 the completeness is low due to roll angle error, while for C12  it is due to extra targets proposed in the broken CCD module. C13 also has slightly lower completeness due to
extra targets proposed in the broken CCD module. C17 shows a two-step fall in the fraction, the cause of which is not clear.
\label{fig:proposed_completeness}}
\end{figure}

\begin{figure*}[tb]
\centering \includegraphics[width=0.99\textwidth]{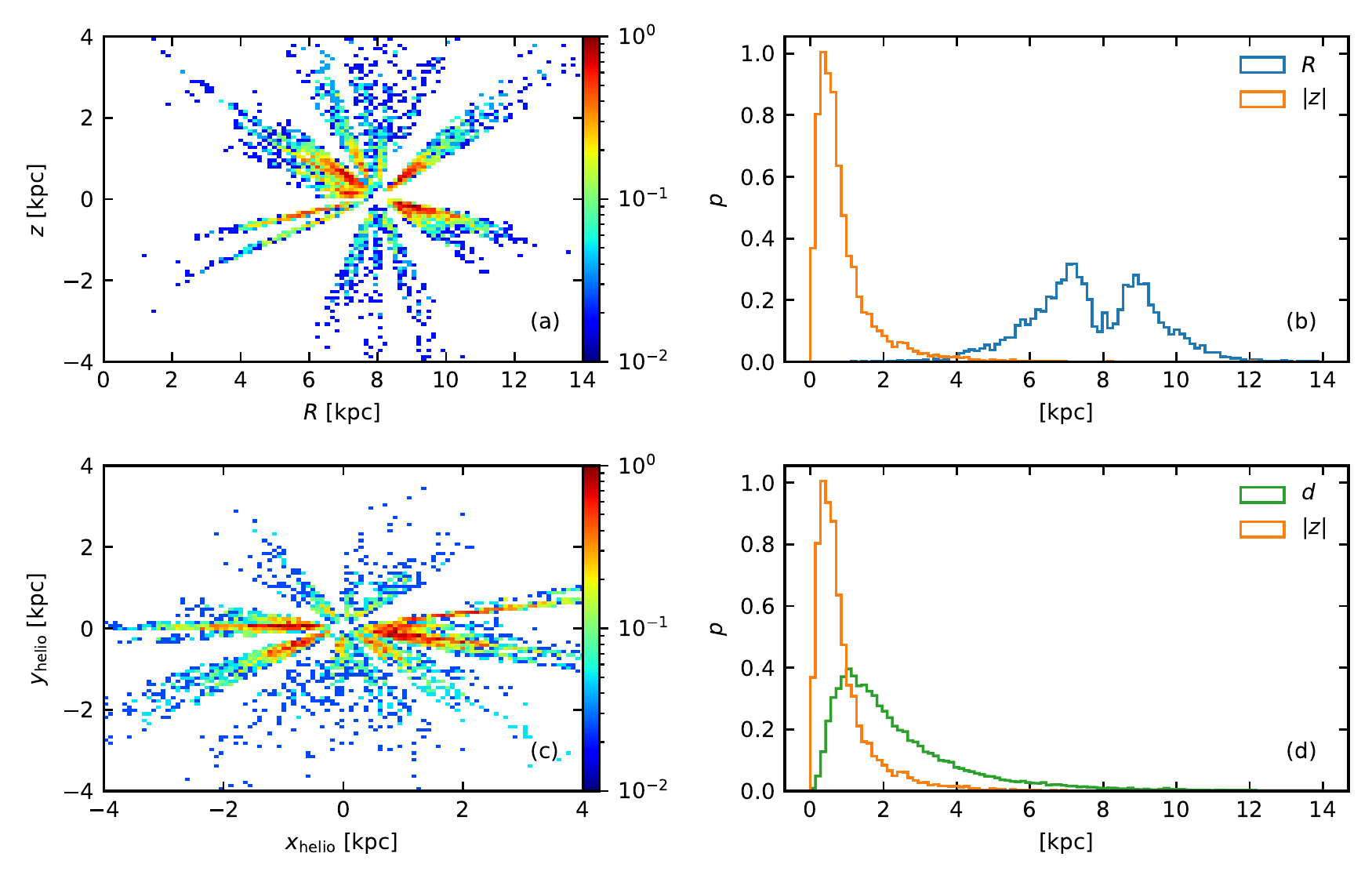}\caption{
The spatial distribution of oscillating giants using
parallax information from {\it Gaia}.
Panels (a) and (b) show the distribution in  Galactocentric cylindrical coordinates $R$ and $z$.
Away from the plane the radial coverage in $R$ extends all the way from 2 kpc to 14 kpc. However, close to the plane
the radial extent is quite limited.
Panels (c) and (d) show the distribution in heliocentric Cartesian coordinates $x_{\rm helio}$ and
$y_{\rm helio}$ and distance $d$.
\label{fig:Rz_dist}}
\end{figure*}

\subsection{Proposed targets}\label{sec:proposed_targets}
We now describe the procedure adopted for creating the proposed target list.
Separate target lists were created for each campaign.  In general, stars were selected to have good quality photometry as defined in \autoref{tab:qualsel}.
Targets in each list were sorted based on priority. A list of various priority classes and their order is given in \autoref{tab:priority}.

Classes with order less than 10, were
for stars known to be giants from either spectroscopy or asteroseismology or were special targets. We included stars with existing spectroscopy from surveys such as APOGEE, RAVE, GAIA-ESO and  SEGUE.
From previous experience with \kepler, it was known that seismic detections were possible for $1.9<\log g< 3.5$. Given spectrocopic $\log g$ have uncertainties of the order of 0.1 dex, a criterion of $\log g < 3.8$ was adopted for the spectroscopic sample. A lower limit on $\log g$  was not adopted as there are very few stars with $\log g < 1.9$.

Classes with order number greater than or equal to 10, were for stars selected based on photometry and they make up most of the proposed sample.
The class with order number 10 was used in all campaigns and provided the bulk of the K2GAP targets.  Stars were selected simply based on color and magnitude and were priority sorted by magnitude. The class with order number 11 makes use of color
\be
l_{\rm color}&=&-0.436 u_{\rm SDSS}+1.129 g_{\rm SDSS}-0.119 r_{\rm SDSS} -0.574 i_{\rm SDSS}+\nonumber\\
&&0.1984,
\ee
defined in term of $u, g, r$ and $i$ band SDSS magnitude.
The classes with order numbers 12 and 13, additionally made use of reduced proper motion
\be
\mu_{\rm reduced}&=&5\log\left(\frac{\sqrt{\mu_{\rm RA}^2+\mu_{\rm Dec}^2}}{\rm arcsec}\right)+r_{\rm SDSS},
\ee
defined in term of proper motion
$\mu_{\rm RA}$ and $\mu_{\rm Dec}$ and $r$ band
SDSS magnitude.

In general, stars were proposed over the entire field of view (see \autoref{sec:fov}). However, for certain dense-field campaigns that look into the plane of the Galactic disc, stars were proposed in one degree circles located at the center of CCD modules. This was done to make the spectroscopic followup easier.

The final target list uploaded to the spacecraft for observation was prepared by NASA and it contained targets from different proposals. Hence, some of our proposed targets were serendipitous selections; their selection was based on other proposals and they just happened to also be on our target list. The serendipitous targets therefore do not follow our proposed priority order.
Guided by \autoref{fig:proposed_completeness} we devised the procedure to exclude these targets. \autoref{fig:proposed_completeness} shows the average fraction of stars selected from our list as a function of our list row number. We see that the fraction is high and almost constant for small row numbers, but it abruptly falls to a low value and remains low for the rest of the list. The sharp fall in fraction marks the special row number up to which the sample selection is complete but beyond which there are serendipitous targets.
Identifying the special row number up to which the sample selection is complete is a change point detection problem. We propose a novel algorithm for quantifying this. The change point is given by
\be
{\rm argmax}\left(\sum_{k=1}^{i} S_i- i f_c  \right),
\ee
\noindent where $S_i$ is a binary state variable, which is 1 if the $i$-th star is selected but otherwise 0. The procedure identifies the row number where the completeness $f$ changes from greater than $f_c$ to less than $f_c$.  Here, $f_c$ is a free
tunable parameter and we set it to 0.8. The actual change point is
not too sensitive to the exact choice of $f_c$.
Using the above procedure, a total of 110791 were found to be free of serendipitous targets, which we refer to as our K2GAP complete sample. From this sample we identify the
101420 stars that are color magnitude complete following  the selection function listed in \autoref{tab:selfunc}.

\subsection{Asteroseismic parameters of giants}
In this paper we use two seismic quantities: the frequency of maximum oscillation power, $\nu_{\rm max}$, and the frequency separation between overtone oscillation modes, $\Delta \nu$. The values we use were the so-called SYD results that are also adopted by Reyes et al. (submitted) and the K2GAP DR3 catalog \citep{2021arXiv210805455Z}.
We used the associate detection probabilities $p_{\nu_{\rm max}}$, based on \citet{2018ApJ...859...64H} and $p_{\Delta{\nu}}$, based on Reyes et al submitted).
A target with $p_{\Delta{\nu}}<0.5$ was assumed to have an invalid $\Delta{\nu}$.  A target with both $p_{\nu_{\rm max}}<0.5$ and $p_{\Delta{\nu}}<0.5$
was assumed to have an invalid $\nu_{\rm max}$. After applying these definitions, a total of 30923 targets had a valid
$\nu_{\rm max}$ and of these, a total of 20708 targets had a valid $\Delta{\nu}$. Note, some stars were observed in multiple campaigns, which means there can be multiple targets (and hence results) that correspond to the same star.
To be consistent with \citet{2016ApJ...822...15S}, we use the SYD results from \citet{2013ApJ...765L..41S} when comparing with \kepler, which had a total of 12919 targets with valid $\nu_{\rm max}$ and $\Delta{\nu}$.

\section{Results}
\subsection{Spatial distribution of oscillating giants}
One  of the main advantages of the K2 mission over the {\it Kepler} mission  is its wider coverage of the Galaxy. This can be seen in \autoref{fig:Rz_dist}, which shows the spatial distribution of oscillating giants in the Galaxy.
The K2 giants span a wide region in the $(R,z)$ plane, while {\it Kepler} giants were confined to a small
range around $R=R_{\odot}$. This is specially useful for
studying the formation and evolution of the Galaxy.
Away from the plane the radial coverage in $R$ extends all the way from 2 kpc to 14 kpc (\autoref{fig:Rz_dist}b, blue curve). However, close to the plane
the radial extent is quite limited (\autoref{fig:Rz_dist}a, $z\sim 0\,$kpc).
In the heliocentric $(x,y)$ projection the giants can be seen in all four quadrants, but there are more of them in the lower half defined by $y_{\rm helio}<0$ (\autoref{fig:Rz_dist}c).
Although the giants extend up to a distance of about 6 kpc
from the Sun, most of the stars are within a distance of about 2.5 kpc (\autoref{fig:Rz_dist}d, green curve).

\begin{figure}[tb]
\centering \includegraphics[width=0.49\textwidth]{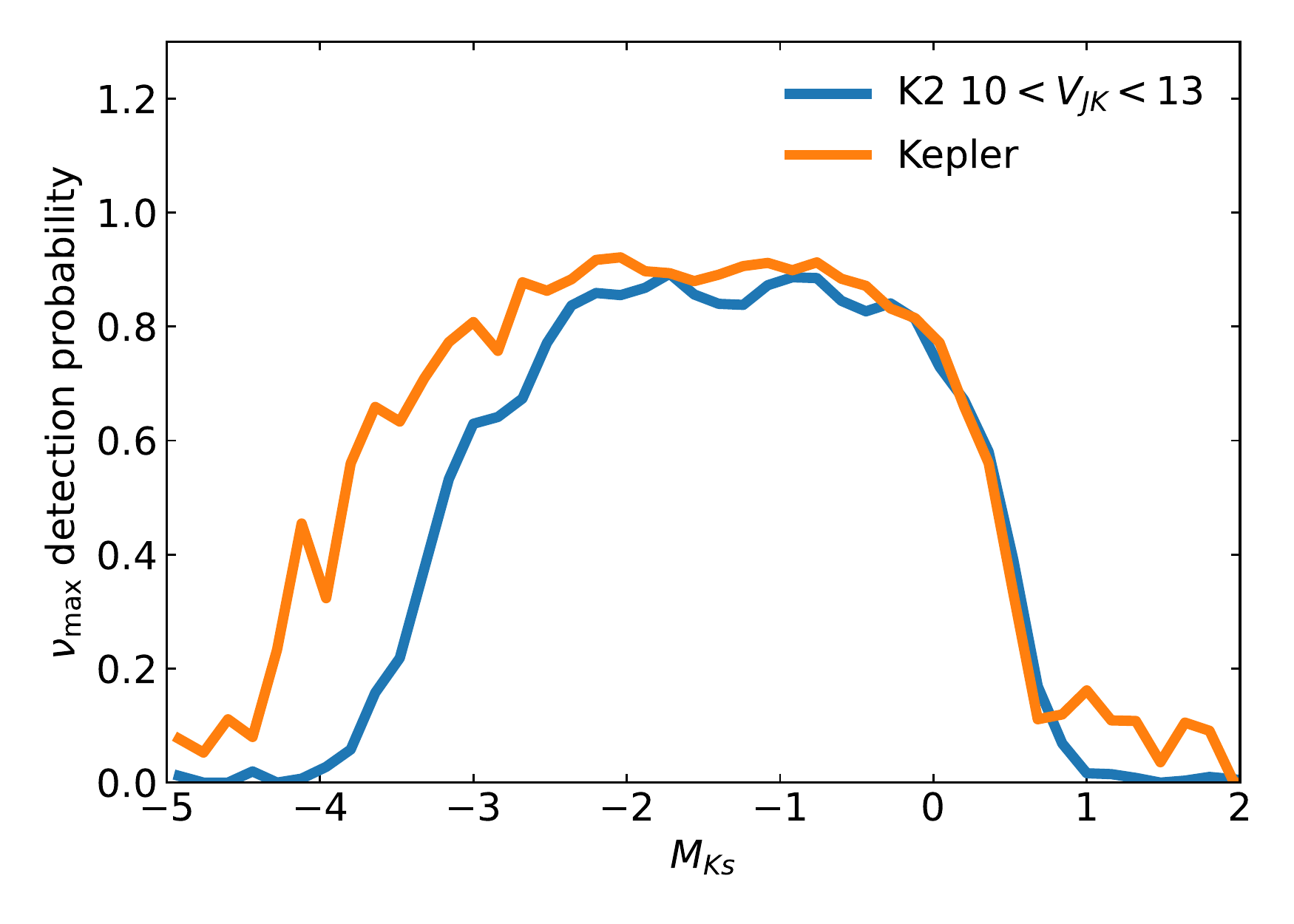}\caption{The probability of detecting $\nu_{\rm max}$ as function of absolute magnitude $M_{Ks}$. The probability is measured as the ratio of stars with $\nu_{\rm max}$ detections to the number of stars observed in each bin of $M_{Ks}$. The average probability in the range
$-2.5<M_{Ks}<0$ is 0.86 for K2 and 0.89 for {\it Kepler}.
\label{fig:completeness_numax_mks}}
\end{figure}

\begin{figure}[tb]
\centering \includegraphics[width=0.49\textwidth]{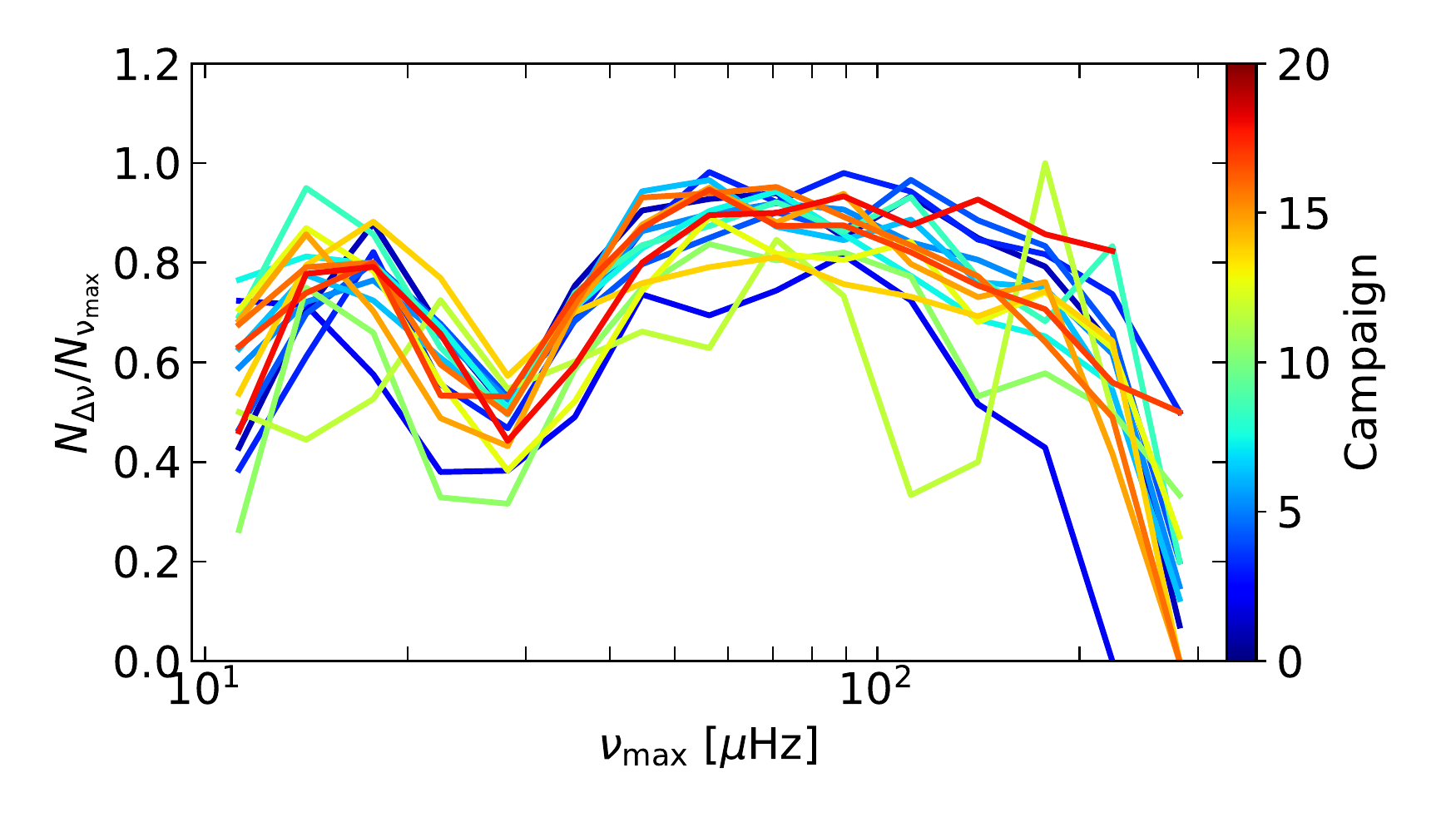}\caption{Fraction of stars with $\nu_{\rm max}$ detections that also have $\Delta \nu$ detections.
\label{fig:dnu_fraction}}
\end{figure}

\subsubsection{Probability to detect $\nu_{\rm max}$.}
\label{sec:numax_prob}
For bright stars the K2 mission is expected to detect oscillations in stars with
$10\lesssim\nu_{\rm max}/\mu{\rm Hz}\lesssim270$.
The lower limit $\nu_{\rm max}$ is due to the duration of the observations and the upper limit is due to the 30 min cadence of the data.
The absolute magnitude, for example $M_{Ks}$,  increases with increasing $\nu_{\rm max}$.
Hence, stars with $\nu_{\rm max}$
measurements should be confined to a range in $M_{Ks}$. This suggests that $M_{Ks}$ can be used to estimate the overall detection probability and
this can be seen in \autoref{fig:completeness_numax_mks}, where we plot the ratio of stars with
$\nu_{\rm max}$ measurement to the number of observed stars as function of $M_{Ks}$.
The probability in the range $-2.5<M_{Ks}<0$
is approximately constant but falls off at either end. Stars with $M_{Ks}>0$ have $\nu_{\rm max}$ that is too high to be measurable while stars with $M_{Ks}<-2.5$ have $\nu_{\rm max}$ that is too low. The average probability in the range
$-2.5<M_{Ks}<0$ was found to be 0.86 for K2 and 0.89 for {\it Kepler}. This represents the overall probability to
detect $\nu_{\rm max}$. The probability for
{\it Kepler} is slightly higher, most likely due
to higher SNR resulting from light curves being
about 12 times longer.
For {\it Kepler} the high detection probability
zone extends to lower values of $M_{Ks}$ ($-3$)
than for K2 ($-2.5$). This is expected due to
{\it Kepler} light curves being
significantly longer than for K2.
For K2, we restrict our analysis to bright stars,
$10<V_{JK}<13$, this is because
fainter stars have low S/N,
which progressively makes it harder to detect $\nu_{\rm max}$, specially for stars with high $\nu_{\rm max}$, which have lower oscillation amplitudes. Lowering the
bright limit on $V_{JK}$ was found to have
no effect on the detection probability profile
shown in \autoref{fig:completeness_numax_mks}.

\subsubsection{Probability to detect $\Delta \nu$ for a given $\nu_{\rm max}$.}
In \autoref{fig:dnu_fraction} we show the
probability of detecting $\Delta \nu$ given a $\nu_{\rm max}$ detection, with each campaign shown separately.
The probability is computed as the fraction of stars with
$\nu_{\rm max}$ detection that also have a $\Delta \nu$ detection.
The fraction shows an undulating behaviour with two peaks, which arise from a global smooth hill shape that peaks at $\nu_{\rm max}\sim60\,\mu$Hz and a local dip near $\nu_{\rm max}$ of 30$\mu$Hz. For the global shape, the drop towards lower $\nu_{\rm max}$ is due to the
lower limit set by the 80 day duration of the light curve; $\Delta \nu$ becomes harder to resolve.
The drop towards higher $\nu_{\rm max}$ is because $\nu_{\rm max}$ approaches the Nyquist frequency. The dip at $\nu_{\rm max} \sim 30\,\mu$Hz corresponds to the location of the red clump (RC) stars. It shows that $\Delta \nu$ is generally harder to detect in clump stars. The asteroseismic completeness for other pipeline results and as a function of mass and radius is considered in \citet{2021arXiv210805455Z}.

\begin{figure*}[tb]
\centering \includegraphics[width=0.99\textwidth]{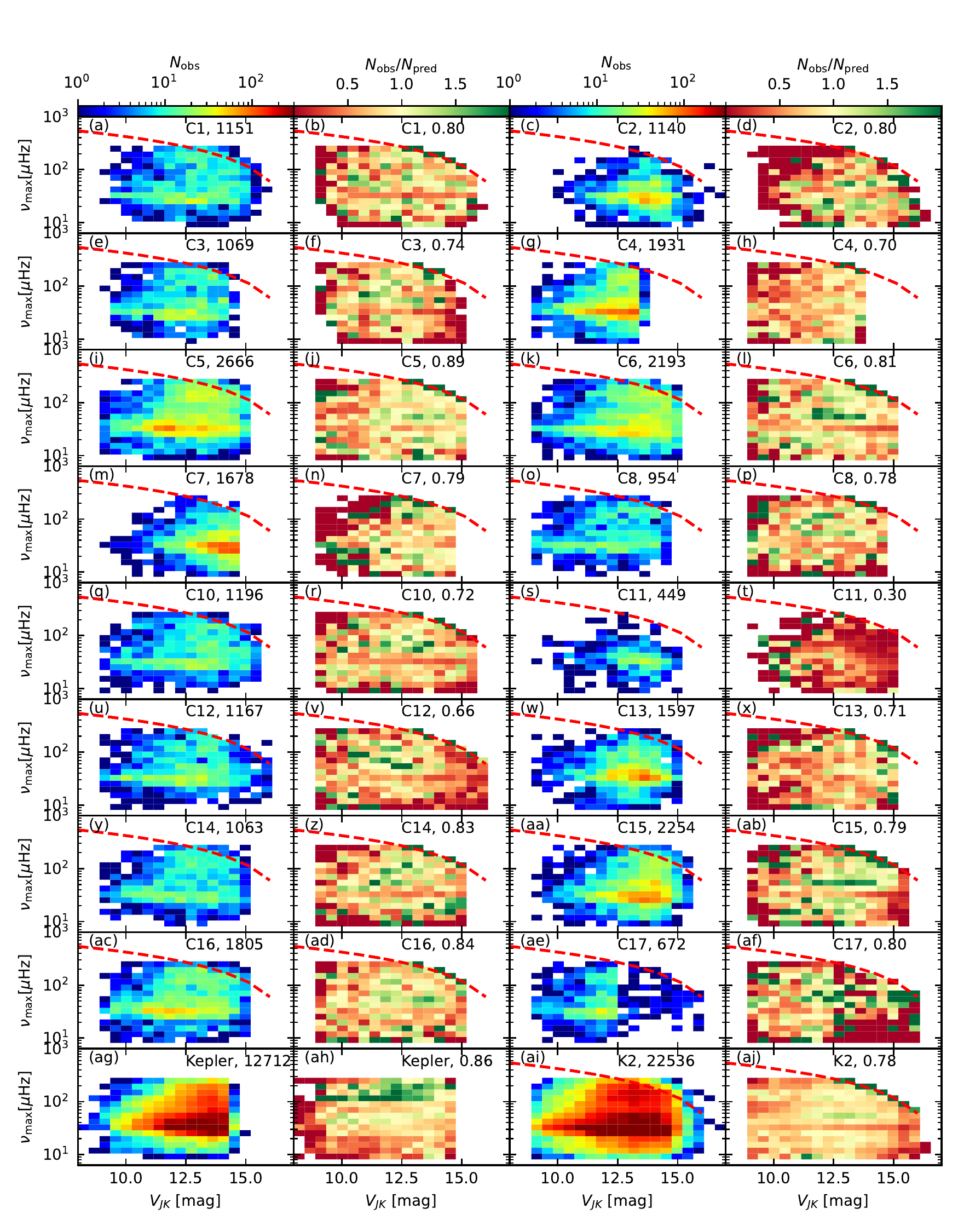}\caption{Distribution of observed (the first and  the third columns) stars in the $(\nu_{\rm max},V_{JK})$ plane for different K2 campaigns. The results for \kepler{} is shown in panel (ag),
while the combined results from all K2 campaigns (except C11) is shown in (ai).
The panels in the second and the fourth columns plot the ratio of observed to predicted oscillating giants in each bin, with the average  recovery rate shown in the top right of each panel. The predictions are based on simulations using {\sl Galaxia}.
The dashed line represents the equation $\nu_{\rm max}=-60(V_{JK}-17)$.The upper right region (above the dashed line) indicates where we cannot detect oscillations due to too low signal-to-noise.
\label{fig:vmag_numax_all}}
\end{figure*}

\subsection{Comparing observed asteroseismic data with Galactic model predictions}
One of the main aims of the K2GAP is to study
the formation and evolution of the Milky Way.
A important step in this process
is to compare the observed asteroseismic data with the predictions of a current state-of-the-art Galactic model. This will help identify any major issues
that need to be addressed before fine tuning the model.
Each K2 campaign was unique with characteristics such as
the light curve duration,
the pointing accuracy, and the crowding varying from campaign to campaign. Hence, it is
important to highlight if any campaigns are in some way problematic. For example, if observations agree with model predictions in some campaigns but not in others,
then that is strong evidence for problematic campaigns.
Having described the selection function and detection completeness we now proceed to performing this model comparison.

We begin by creating a synthetic catalog of stars in accordance with a Galactic model that satisfy the same selection function as the observed data, which is available for download \footnote{\url{http://www.physics.usyd.edu.au/k2gap/download/k2\_observed\_simulated.fits}}. Next, we
compare the predicted and observed distribution of various stellar properties, such as apparent magnitude  $V_{JK}$, $\nu_{\rm max}$, and $\kappa_M$; the latter being a seismic mass proxy that includes both $\nu_{\rm max}$ and $\Delta \nu$.

To sample data from a prescribed Galactic model we use the {\sl Galaxia}\footnote{\url{http://galaxia.sourceforge.net}} code
\citep{2011ApJ...730....3S}. It uses a Galactic model
that is initially based on the {\sl Besan\c{c}on} model
by \citet{2003A&A...409..523R} but with some crucial modifications, which are described in \citet{2019MNRAS.490.5335S}. One of the most significant  change was the shift in mean effective metallicity of the thick disc from $-0.78$ to $-0.162$. This model
with a metal rich thick disc is hereafter referred to as Galaxia(MR).

The main steps in creating the synthetic catalog are
as follows. Using {\sl Galaxia}, we create magnitude limited samples with $J<15$ for each campaign over a circular area of 8 degree radius.
The stars are then filtered in accordance with  the selection function
on color, magnitude, and angular coordinates
as described in \autoref{tab:selfunc}. We provide
a python function\footnote{\href{https://github.com/sanjibs/k2gap}{https://github.com/sanjibs/k2gap}} to do this.
Next, the synthetic stars are sub-sampled (without
replacement) to match the total number of
observed stars in each campaign that follow the selection function (column 5 in \autoref{tab:selfunc}).
In reality, to reduce Poisson noise we over-sample
the synthetic stars by a factor of 10.
For reference purposes we also show results
for the {\it Kepler} mission. The selection function
that we adopt for {\it Kepler}  is given by Equations (2) and (3) of \citet{2019MNRAS.490.5335S}. Full details to
reproduce the {\it Kepler} selection function can be found in \citet{2016ApJ...822...15S}.

The seismic quantities $\nu_{\rm max}$ and $\Delta \nu$ for the synthetic stars are estimated from effective temperature, $T_{\rm eff}$, surface gravity, $g$, and density, $\rho$, using the following asteroseismic scaling relations \citep{1991ApJ...368..599B,1995A&A...293...87K,1986ApJ...306L..37U}.
\be
\frac{\nu_{\rm max}}{\nu_{\rm max,\odot}}=\frac{g}{g_{\odot}}\left(\frac{T_{\rm eff}}{T_{\rm eff,\odot}}\right)^{-0.5} \label{equ:scaling_numax} \text{\ and\ \ }
\frac{\Delta \nu}{\Delta \nu_{\odot}}=f_{\Delta \nu}\left(\frac{\rho}{\rho_{\odot}}\right)^{0.5}. \label{equ:scaling_dnu}
\ee
Here, $f_{\Delta \nu}$
is the correction factor
derived by \citet{2016ApJ...822...15S} by analyzing theoretical oscillation frequencies with GYRE \citep{2013MNRAS.435.3406T} for stellar
models generated with MESA \citep{2011ApJS..192....3P,2013ApJS..208....4P}. We used the code ASFGRID\footnote{\href{http://www.physics.usyd.edu.au/K2GAP/Asfgrid}{http://www.physics.usyd.edu.au/K2GAP/Asfgrid}} \citep{2016ApJ...822...15S}
that computes the correction factor as a function of metallicity $Z$, initial
mass $M$, evolutionary state $E_{\rm state}$ (pre or post helium ignition), $T_{\rm eff}$, and $\log g$.
A random scatter based on the median observed uncertainty
for each campaign is added to estimated values of $\nu_{\rm max}$ and $\Delta \nu$.

As discussed earlier in \autoref{sec:numax_prob},
we expect to detect oscillations only in the range
$10\lesssim\nu_{\rm max}/\mu{\rm Hz}\lesssim 270$.
However the probability to measure $\nu_{\rm max}$
is not constant over this range. The oscillation amplitude in general decreases with $\nu_{\max}$ and the overall noise in the extracted power spectrum increases with increasing apparent magnitude. This means that at fainter magnitudes the high $\nu_{\max}$ stars are less likely to show detectable oscillations.
To model the seismic detection probability we follow
the scheme presented by \citet{2011ApJ...732...54C} and \citet{2016ApJ...830..138C}. The exact procedure
that was adopted is given in Section 3.4 of \citet{2019MNRAS.490.5335S}. In short, the oscillation
amplitude was estimated based on stellar luminosity, mass, and temperature. The apparent magnitude was used to compute the instrumental photon-limited noise in the power spectrum, which when combined with granulation noise gave the total noise.
The mean oscillation power and the total noise were then used to derive the probability of detecting oscillations (with less than 1\% possibility of false alarm). Stars with a detection probability of greater than 0.9 were assumed to be detectable.

\subsubsection{The distribution of stars in the $(\nu_{\rm max}, V_{JK})$ plane} \label{sec:numax_vmag_dist}
\autoref{fig:vmag_numax_all} shows the distribution of
observed stars in the $(V_{JK},\nu_{\rm max})$ plane
for different K2 campaigns (first and third columns).
Results from \kepler{} (panel (ag)) and the combination all K2 campaigns (panel (ai)) are also shown.
The heat maps in the second and fourth columns show the number ratio of observed to predicted giants.
It can be see that $\nu_{\rm max}$ can be measured
for stars as faint as $V_{JK}=16$. However, the
efficiency seems to decrease beyond $V_{JK}=16$
(dark region in \autoref{fig:vmag_numax_all}aj).
It can be seen that in K2 there is a lack of observed stars in the upper right corner (above the red dashed line). The red dashed line roughly
identifies the threshold beyond which the theory predicts that the
oscillations would be hard to detect because of too low oscillation amplitude (high \numax) and to high noise (faint stars).

The ratio of observed to predicted number of stars
varies from campaign to campaign, and is typically
between 0.66 and 0.89. When averaged over all campaigns
the ratio is 0.78. From \autoref{sec:numax_prob} (\autoref{fig:dnu_fraction}) we know that 14\% of giants are expected to be missing. The extra 8\% of missing
giants are most likely due to the fact that the model overpredicts
the number of giants (stars with $M_{Ks}<2$) compared to dwarfs for stars with $J-Ks>0.5$.
This needs further investigation.
It could also be due
to the turnoff stars being bluer in the model, and hence getting excluded in our $J-Ks>0.5$ cut used to select the stars.

Campaign C11 has unusually low numbers of observed stars
compared to the prediction, with a ratio of 0.3.
This could be related to the data quality,
which is doubtful for a couple of reasons.
Firstly, due to an error in the initial roll angle,
a correction was applied 23 days
into the campaign and as a result the campaign
was split into two segments.
Secondly, C11, which was pointing towards the Galactic bulge, has the highest stellar density amongst all the campaigns analyzed here. Due to the high stellar density the pipelines used for extracting light curves as well as for the subsequent asteroseismic analysis are most likely operating outside their nominal design range. This can severely hamper the
quality of the derived asteroseismic parameters.
In the following two sections we will look at the distributions shown in \autoref{fig:vmag_numax_all} collapsed onto either axis.

\begin{figure}[tb]
\centering \includegraphics[width=0.49\textwidth]{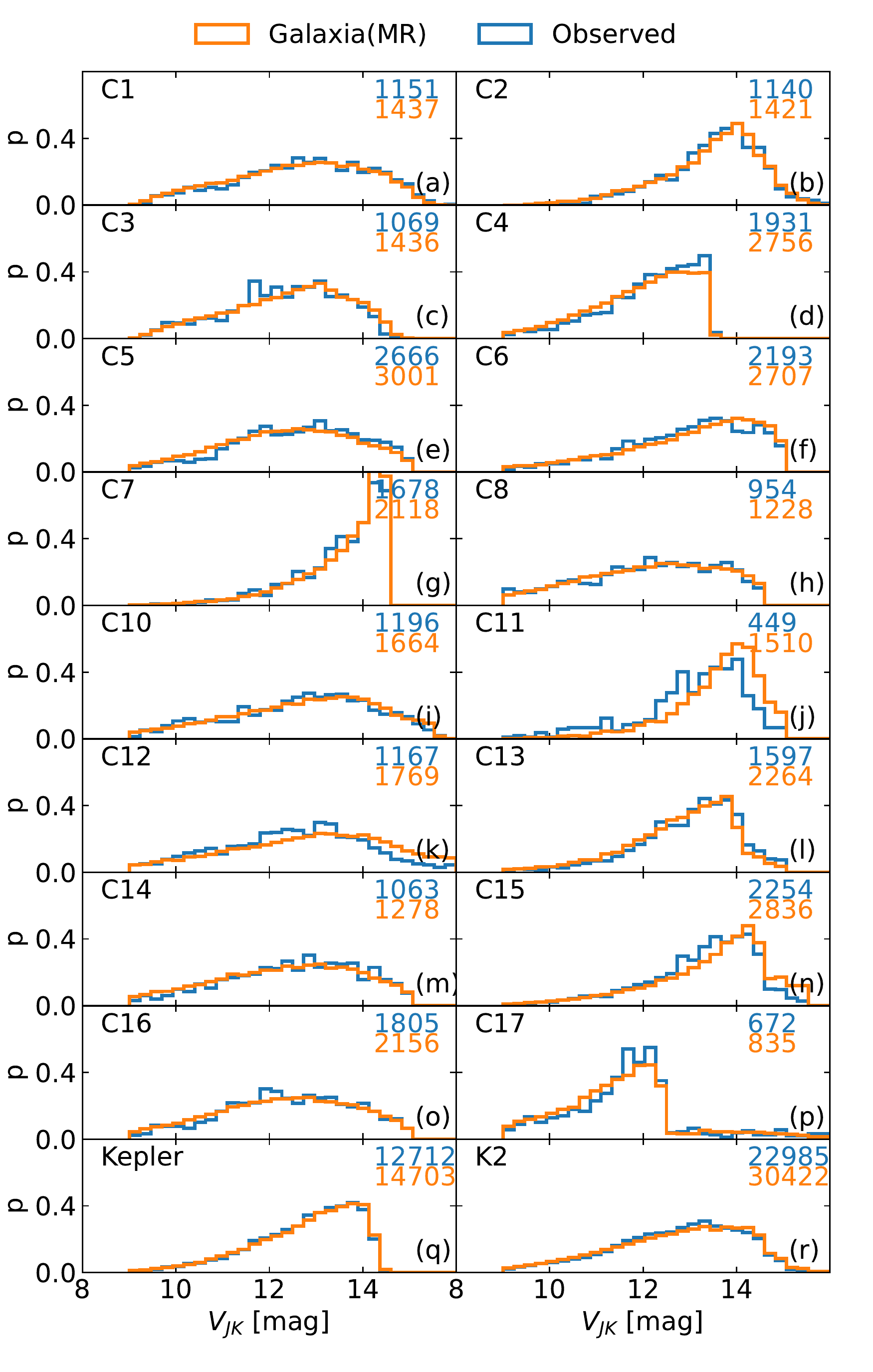}\caption{Magnitude distribution of observed oscillating giants from K2 along with predictions from {\sl Galaxia} (model MR). The number of stars with $\nu_{\rm max}$ detections in the observed sample and those predicted by the model are also listed in each panel.
\label{fig:vmag_dist}}
\end{figure}

\subsubsection{The distribution of apparent magnitude}
In \autoref{fig:vmag_dist} we show the distribution
of apparent magnitude $V_{JK}$ of the giants with
$\nu_{\rm max}$ detections, which shows good agreement between the model and the data. The good match is primarily a reflection
of the fact that the model correctly reproduces
the number of giants as a function of magnitude.
A good match in $V_{JK}$ is in some sense a necessary
condition before we
embark on a more detailed comparison of model
predictions with observations.
The number of observed and
predicted giants is listed on each panel.
It can be seen that the model overpredicts
the number of oscillating giants, which we discuss
in more detail later on.

\begin{figure}[tb]
\centering \includegraphics[width=0.49\textwidth]{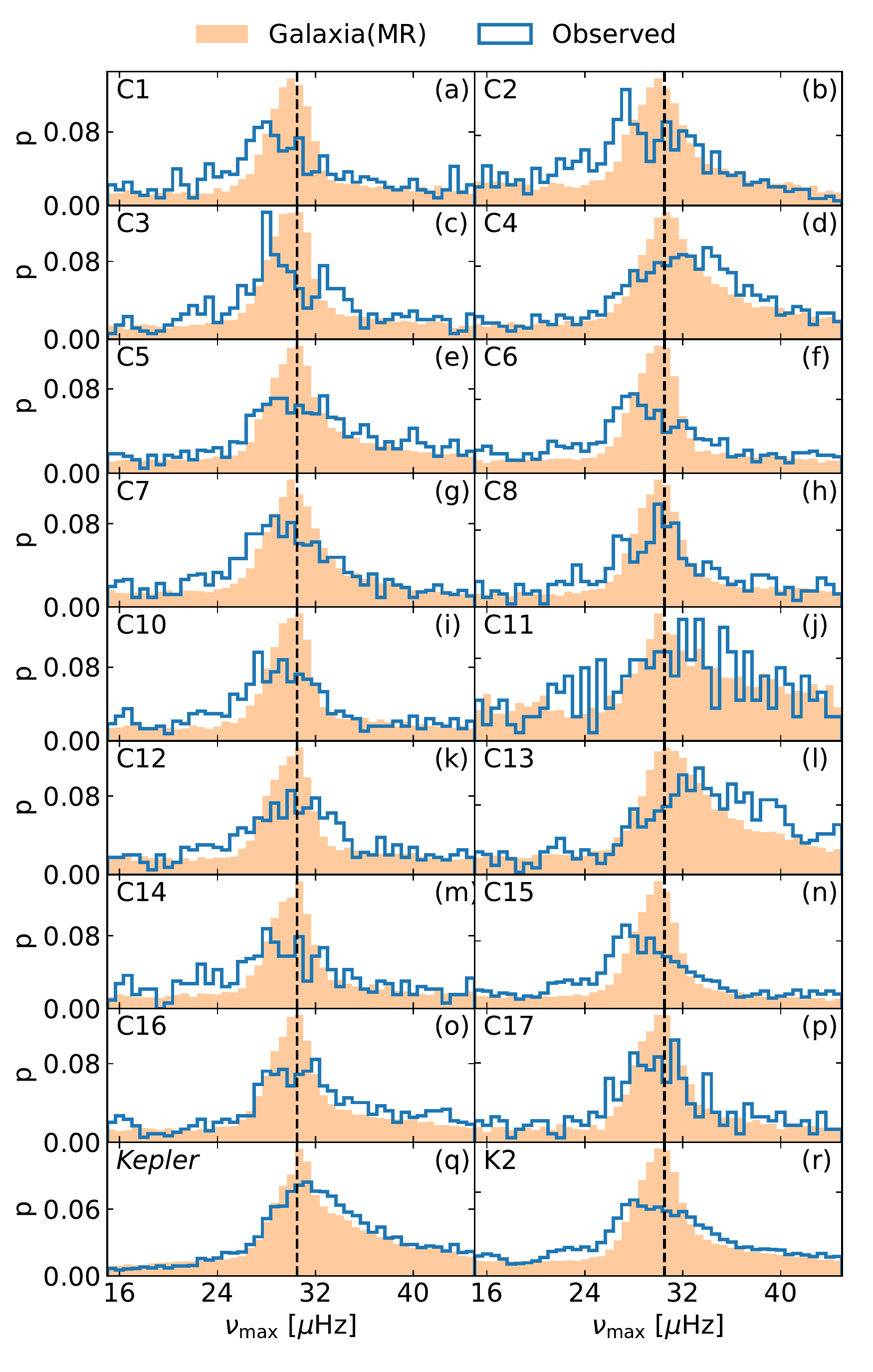}\caption{The probability distribution of $\nu_{\rm max}$ for observed and predicted oscillating giants.  The dashed line, $\nu_{\rm max}=30.5\ \mu$Hz, shows the approximate location of the peak in the distribution of the predicted stars. The peak corresponds to the location of the RC giants. Panel (q) shows the same distributions for Kepler. Panel (r) shows the combined
data from all K2 campaigns.
The observed peak is systematically lower compared to the predictions, except for C4 and C13, which point towards the anti-center direction.
\label{fig:numax2_dist}}
\end{figure}

\subsubsection{The distribution of $\nu_{\rm max}$}
In \autoref{fig:numax2_dist} we now show the observed distribution of $\nu_{\rm max}$ alongside the model predictions.
The distribution shows a peak at around $\nu_{\rm max}$
of 30.5 $\mu$Hz, corresponding to the RC stars.
For K2, the peak of observed stars is significantly  shallower compared to the prediction.
This could be due to the uncertainty in $\nu_{\rm max}$
being underestimated in the model or because the
model is inaccurate. It could also be due
to RC stars preferentially `escaping detection'.
For K2, the observed peak is systematically shifted to lower $\nu_{\rm max}$ compared to the predictions,   except for C4 and C13. Interestingly,
both C4 and C13 point towards the anti-center
direction suggesting the model requires some changes.

\begin{figure*}[tb]
\centering \includegraphics[width=0.99\textwidth]{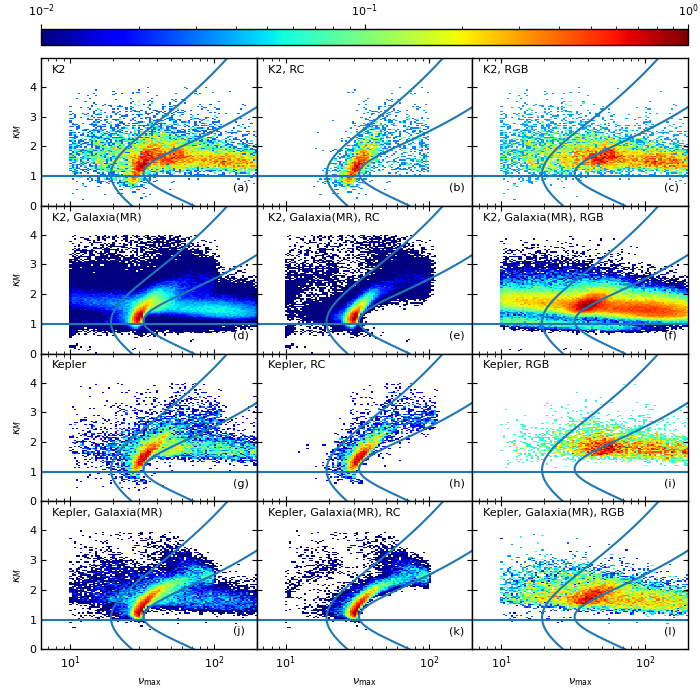}\caption{Distribution of stars in the $(\kappa_M,\nu_{\rm max})$ plane. Panels(a,b,c) show result for K2 stars. Panels(d,e,f) show theoretical predictions for K2. Panels(g,h,i) show results for \kepler{} stars. Panels(j,k,l) show theoretical predictions for \kepler.
Left columns show the results for all stars,
the middle columns show the results for RC stars while the right columns show the results for RGB stars.
The solid lines (two curves and a horizontal line) are drawn to aid the eye when comparing the distributions in the different panels.
The two curves are from  \citep{2019MNRAS.490.5335S}, which were designed to roughly identify the RC stars.
\label{fig:numax_kappa_m_combined}}
\end{figure*}

\begin{figure*}[tb]
\centering \includegraphics[width=0.99\textwidth]{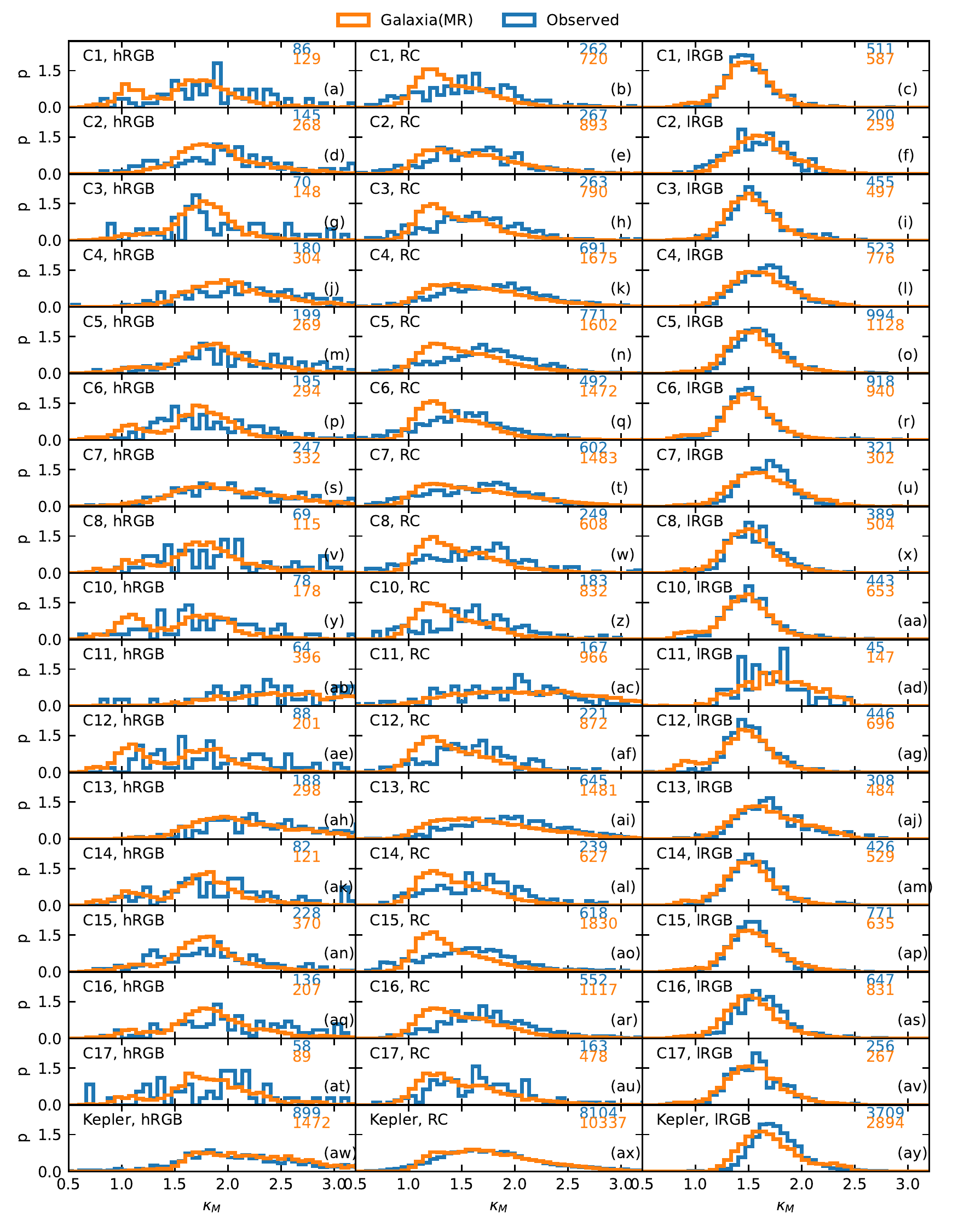}\caption{ The distribution of $\kappa_M$ for different K2 campaigns and \kepler{}. Distributions for three different stellar classes, high-luminosity RGB, RC, and low luminosity RGB stars are shown separately. The classification was done using the two solid curves in \autoref{fig:numax_kappa_m_combined}.
In each panel, the number of observed and predicted stars are listed on the right hand side. The bottom row shows the distributions for \kepler{}.
\label{fig:kappa_m}}
\end{figure*}

\begin{figure*}[tb]
\centering \includegraphics[width=0.99\textwidth]{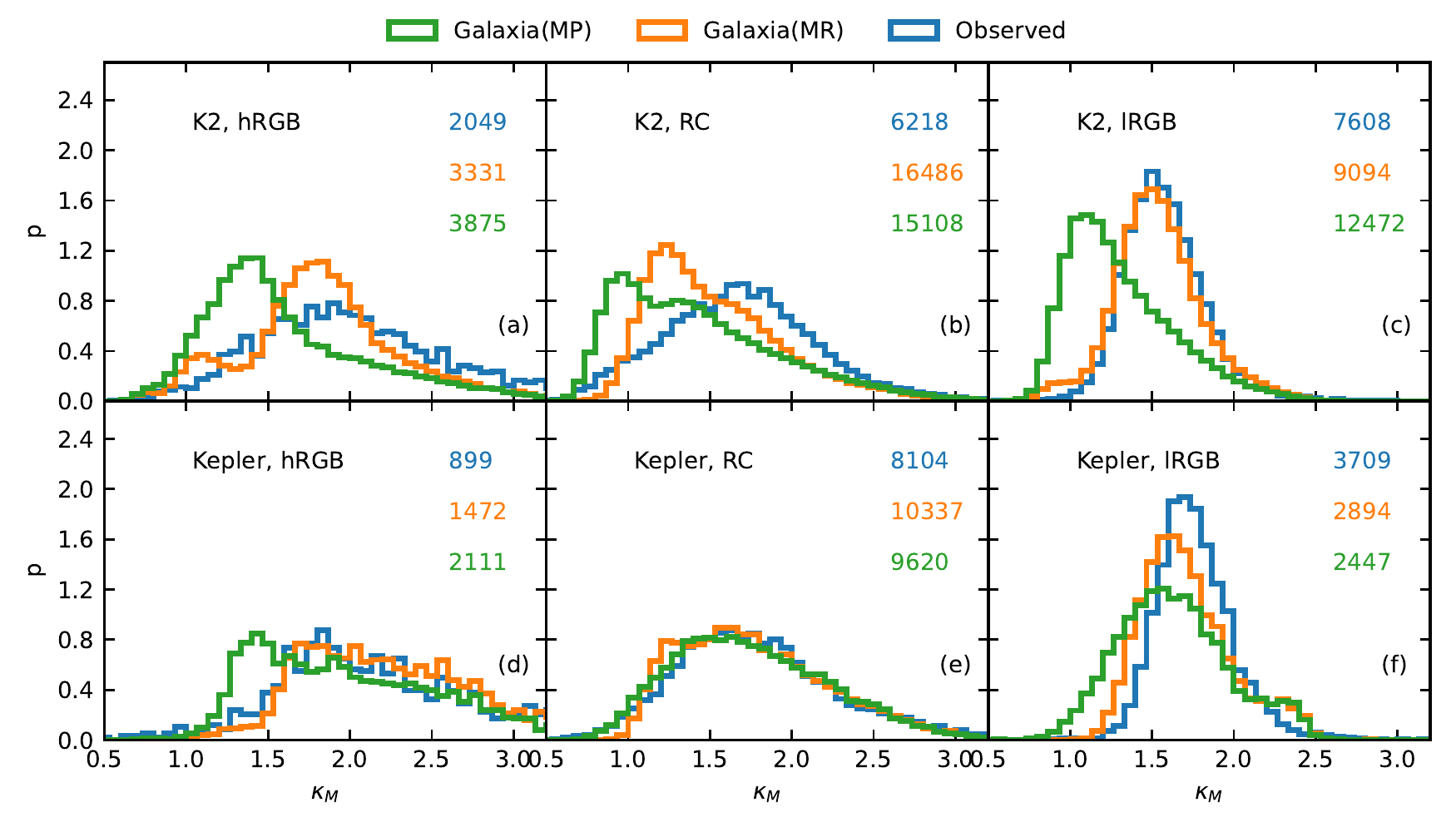}\caption{The distribution of $\kappa_M$ for oscillating giants in {\it Kepler} and K2. Distributions for high-luminosity RGB, RC, and low luminosity RGB stars are shown separately. Predictions based on {\sl Galaxia}
for the metal poor (MP) and metal rich (MR) model are also shown in each panel.
\label{fig:dist_kappa_m_combined}}
\end{figure*}

\subsubsection{The distribution of stars in the $(\nu_{\rm max}, \kappa_M)$ plane}
We now turn to
\be
\label{equ:scaling_m1}
\kappa_{M} &  = &  \left(\frac{\nu_{\rm
max}}{\nu_{\rm max,
\odot}}\right)^{3}\left(\frac{\Delta \nu}{
\Delta \nu_{\odot}}\right)^{-4},
\ee
which essentially is a temperature-independent seismic proxy for stellar mass.
Mass is one of the most useful asteroseismic quantities because it
helps us to determine the stellar age. Hence, comparing
the observed distribution of $\kappa_M$ with theoretical models is central to the theme of Galactic archaeology. Here, we begin by studying the distribution of stars in the
$(\nu_{\rm max}, \kappa_M)$ plane. This plane is also useful to study the
properties of RC and red giant branch (RGB) stars separately, because as
\citet{2019MNRAS.490.5335S} showed, it can be used to segregate RC from RGB stars.
The RC stars show a very sharp edge in this plane;
a feature that was used by \citet{2021MNRAS.501.3162L} to measure the
intrinsic scatter of asteroseismic scaling relations.

\autoref{fig:numax_kappa_m_combined} shows the $(\nu_{\rm max}, \kappa_M)$ distribution for the K2GAP stars (top row) and for the \kepler{} sample (third row).
The model predictions for K2 and \kepler{} are shown in the second and fourth rows, respectively.
The distributions of RC (middle column) and RGB (right column) stars are also shown separately. The RGB stars are distributed over a wide range of $\nu_{\rm max}$ and $\kappa_M$.
However, RC stars form a diagonal sequence, with most of them lying in a narrow region marked by the two solid curves. These curves were designed by  \citet{2019MNRAS.490.5335S}
to segregate the RC from the RGB. Here, they aid the eye when comparing the distributions in the different panels. The sharp right edge in the RC distribution is very clear. The over density in the RGB distribution is due to the RGB-bump stars,
and it partly overlaps with the location of the majority of the RC stars. The distributions for K2 and \kepler{} are very similar. However, the RC sequence
is slightly sharper for \kepler, most likely due to more precise $\nu_{\rm max}$ and $\kappa_M$. Overall, the model predictions match well with the observed distributions.
A comparison of panel (b) with other panels in the middle column shows that K2 has a significant number of stars to the right of the rightmost solid curve, which is neither predicted by the model (\autoref{fig:numax_kappa_m_combined}e) nor it is present in the \kepler{} data (\autoref{fig:numax_kappa_m_combined}h). This suggests that in K2 some RGB stars are misclassified as RC stars, which is not unexpected given the K2 observations are relatively short making the seismic RC/RGB distinction more uncertain \citep{2018MNRAS.476.3233H}.
The RC stars with high $\nu_{\rm max}$ and high $\kappa_M$ are secondary clump stars. The models (\autoref{fig:numax_kappa_m_combined}e,k) suggest that they have an upper limit of around 100 $\mu$Hz for $\nu_{\rm max}$. In {\it Kepler} we do see these stars all the way till 100 $\mu$Hz. However, the K2
data seems to have relatively fewer of them.
The models also predict a second sequence at low $\nu_{\rm max}$ (stars at the end of helium core burning; the so-called AGB clump stars), but this is not seen in the data.
Finally, in the models, the location of the RGB-bump is offset with respect to that in observations, with the model's location being shifted to lower $\nu_{\rm max}$, which is a known problem of contemporary stellar evolution models \citep{2020A&A...635A.164S}.

\subsubsection{The distribution of stars in $\kappa_M$}
The distribution of $\kappa_M$ is one of the most sensitive tests
of the Galactic models, because it is related to mass, which in turn is related to age. Hence, the distribution of $\kappa_M$ is effectively the age distribution of the stars.
As such, it is sensitive to the star formation history  and
radial migration in the Galaxy. From \citet{2019MNRAS.490.5335S}
we know that for a given age, the mass of a star is correlated with its metallicity. Hence, the distribution of $\kappa_M$ indirectly also
probes the age metallicity relation.
Studies using \kepler{} showed that models over predict the
number of low $\kappa_M$ stars \citet{2016ApJ...822...15S}.
However, doubts about the reproducibility of the complicated selection
function of \kepler{}, prevented us from drawing any strong conclusions.
This problem was alleviated by K2GAP, because of its simple selection function.
Using data from four K2 campaigns, \citet{2019MNRAS.490.5335S}
showed that most of the discrepancy was due to the
metallicity of the thick disc being too low in the models.
In \autoref{fig:kappa_m} we repeat the same analysis but now
using 16 K2 campaigns.
Distributions for high-luminosity RGB (hRGB), RC, and low-luminosity RGB (lRGB) stars are shown separately.
This was done because, for K2 the probability to detect \dnu\ varies with stellar type (\autoref{fig:dnu_fraction}), suggesting systematic
effects for different stellar types.
The stars were split into these categories based on their location in the $(\nu_{\rm max}, \kappa_M)$ plane (see \autoref{fig:numax_kappa_m_combined}); hRGB (left of leftmost solid curve), RC (between the two solid curves) and lRGB (right of rightmost solid curve).
Our results confirm the findings of \citet{2019MNRAS.490.5335S}.
\autoref{fig:kappa_m} shows that for the lRGB stars the observed $\kappa_M$ distribution is in good agreement with  the predictions.
The agreement is also good for hRGB stars, but the models seems to slightly
over predict the number of low mass stars. However, for the RC the
models seem to significantly over predict the number of low mass stars.
This can be seen more clearly in \autoref{fig:dist_kappa_m_combined}
where we combine the results from all K2 campaigns to boost the sample
size. The predicted distribution from the old {\sl Galaxia} model,
with its very metal poor thick disc,
is also shown for comparison. It cannot even reproduce the $\kappa_M$
distribution for the lRGB stars.
A more detailed quantitative comparison is given in \autoref{tab:kappa_m}, where we list the ratio of observed to predicted median $\kappa_M$ for different K2 campaigns, which simply reinforce the qualitative trends discussed above.

The mismatch between the observed and the predicted
distribution for RC stars (\autoref{tab:kappa_m}) could be due to the model being
inaccurate. For example, inaccuracies in the Galactic model (or underlying stellar models)
or inaccuracies in predicting the asteroseismic parameters for
the modelled sample. However, given that K2 fails to measure \dnu\
for a significant number of RC stars (\autoref{fig:dnu_fraction}),
it is possible that low mass stars preferentially
evade \dnu\ measurement. There is some circumstantial
evidence to support this.  For \kepler\ we do not have any
\dnu\ incompleteness and we also do not see any discrepancy with model predictions.

\begin{table}
\caption{Ratio of observed median $\kappa_M$ to that predicted by {\sl Galaxia} (MR) for different giant classes. Uncertainties on the computed ratio are also listed. C11 being an outlier is excluded when averaging over all K2 campaigns.}
\begin{tabular}{llll}
\hline
Campaign  & hRGB & RC & lRGB \\
\hline
1          & $1.15 \pm 0.04$ & $1.14 \pm 0.02$ & $0.999 \pm 0.006$ \\
2          & $1.06 \pm 0.03$ & $1.04 \pm 0.02$ & $0.98 \pm 0.01$ \\
3          & $1.07 \pm 0.05$ & $1.16 \pm 0.02$ & $1.019 \pm 0.007$ \\
4          & $1.09 \pm 0.02$ & $1.11 \pm 0.01$ & $1.033 \pm 0.007$ \\
5          & $1.07 \pm 0.02$ & $1.16 \pm 0.01$ & $1.036 \pm 0.005$ \\
6          & $1.03 \pm 0.02$ & $1.14 \pm 0.01$ & $1.002 \pm 0.005$ \\
7          & $1.02 \pm 0.02$ & $1.08 \pm 0.01$ & $1.025 \pm 0.008$ \\
8          & $1.11 \pm 0.05$ & $1.16 \pm 0.02$ & $1.033 \pm 0.008$ \\
10         & $1.1 \pm 0.04$ & $1.19 \pm 0.03$ & $1.027 \pm 0.007$ \\
11         & $0.82 \pm 0.04$ & $0.94 \pm 0.02$ & $0.95 \pm 0.02$ \\
12         & $1.17 \pm 0.05$ & $1.18 \pm 0.02$ & $1.029 \pm 0.007$ \\
13         & $1.13 \pm 0.02$ & $1.14 \pm 0.01$ & $1.01 \pm 0.01$ \\
14         & $1.07 \pm 0.03$ & $1.19 \pm 0.02$ & $1.022 \pm 0.007$ \\
15         & $1.05 \pm 0.02$ & $1.17 \pm 0.01$ & $1.005 \pm 0.005$ \\
16         & $1.13 \pm 0.03$ & $1.17 \pm 0.01$ & $1.059 \pm 0.006$ \\
17         & $1.09 \pm 0.05$ & $1.16 \pm 0.02$ & $1.01 \pm 0.01$ \\
\hline
K2         & $1.082 \pm 0.009$ & $1.153 \pm 0.004$ & $1.022 \pm 0.002$ \\
Kepler     & $0.96 \pm 0.01$ & $1.018 \pm 0.003$ & $1.041 \pm 0.002$ \\
\hline
\end{tabular}
\label{tab:kappa_m}
\end{table}

\section{Asteroseismic scaling relations and
Galactic Archaeology}
{Asteroseismology can provide ages for giant stars
and hence is a promising tool for studying Galactic structure and evolution. However, it has proven to be difficult to check the accuracy of the ages and masses estimated by asteroseismology, due to the shortage of independent estimates of mass and age. }
Earlier studies indicated that asteroseismology
overestimated masses, found by comparing them with the expected mass of metal poor giants in the {\it Kepler} sample \citep{2014ApJ...785L..28E}, and dynamical mass measurements of binary systems \citep{2016ApJ...832..121G}. Based on \citet{2009MNRAS.400L..80S}, \citet{2016ApJ...822...15S} showed that there are theoretically motivated
corrections to the $\Delta \nu$ scaling relation,
which are important to take into account.
These corrections are enough to resolve
the discrepancy for metal poor giants noticed by
\citet{2014ApJ...785L..28E}.
However, the situation regrading the eclipsing binaries is more complicated. \citet{2016ApJ...832..121G} suggested 15\%
overestimation of mass in spite of
$\Delta \nu$ corrections, while the work by \citet{2018MNRAS.476.3729B}
suggested a good match with the dynamical masses for at least
some binaries.

Population synthesis-based Galactic models, provide an indirect way to validate the asteroseismic estimates, assuming that the models are sufficiently accurate.
These models have been constructed independently of asteroseismology and built to satisfy a number of observations, such as photometric star counts, kinematics, and stellar abundances from spectroscopic surveys.
As mentioned in the previous section, studies using the {\it Kepler} mission have revealed that the models predict too many low mass stars  compared to the observed mass distributions,  both for giants \citep{2016ApJ...822...15S} and subgiants \citep{2017ApJ...835..163S}.
This raised doubts on the accuracy of the asteroseismic scaling relations, the Galactic models,
and/or the selection function.
\citet{2019MNRAS.490.5335S} revisited this  by analyzing asteroseismic data from four campaigns of the K2 mission, which had a well defined selection function. They showed that if the metallicity distribution in the Galactic models is updated to match the measurements from recent spectroscopic surveys, the distribution of asteroseismic masses (for low luminosity giants) is in good agreement with the model predictions.
And as mentioned in the previous section, that result is confirmed here using data from all K2 campaigns.

A number of new interesting developments have happened
since the analysis by \citet{2019MNRAS.490.5335S}.
\citet{2019MNRAS.490.5335S} updated the metallicity in the Galactic model based on results from APOGEE-DR14, but
in the latest  APOGEE-DR16 data release the
metallicity of thick disc stars is lower by about
-0.06 dex. This will again make the models
overpredict the number of low mass stars.

Recent results by \citet{2021MNRAS.tmp.1920S, 2021MNRAS.506.1761S, 2020arXiv201113818S} suggest that the thick disc
model adopted by {\sl Galaxia} is too simplistic.
These results suggest that there is no distinct
thick disc, instead the whole disc is considered to be a continuous sequence of stars in age, with no natural boundary between thick and thin discs. The stellar abundances were shown to be a function only of stellar age and birth radius with the radial migration playing a crucial role in moving stars from their place of birth. The effect of the new
model on the distribution of stellar masses is expected to be small, but it still needs to be tested.

\citet{2021MNRAS.506.1761S} showed that kinematics
can be used to estimate the age of an ensemble of stars and hence test the asteroseismic scaling relations. They concluded that the asteroseismic
ages for {\it Kepler} stars is underestimated by
at least 10\% (or mass overestimated by about 2.6\%). However, no such correction is required for K2.
The exact reason for the systematic
difference between {\it Kepler} and K2 is not yet clear, however, it is clear that this systematic is due to the K2 light curve being significantly shorter.
\citet{2021arXiv210805455Z} demonstrate in their Figure 5 that {\it Kepler} data, when shortened to K2 time baselines, leads to an underestimation of $\nu_{\rm max}$ by about 1\% (hence mass by 3\%), which is in good agreement with findings of  \citet{2021MNRAS.506.1761S}.
This is consistent with our results for lRGB stars as shown in \autoref{tab:kappa_m}, which shows that for K2, the mass is overestimated by 2.2\% while for {\it Kepler} it is 4.1\%. In other words, the K2-based masses are about 1.9\% lower than for {\it Kepler}.
Another interesting systematic associated with asteroseismic mass
and age is given by \citet{2021AJ....161..100W},
who suggest that the ages of high $[\alpha/{\rm Fe}]$ stars
are underestimated by stellar models by about 10\%, in excellent agreement with \citet{2021MNRAS.506.1761S}.
Traditionally, the \citet{2005essp.book.....S} formula is
used to account for $[\alpha/{\rm Fe}]$ enhancement
by assuming solar composition, but increasing the
metallicity. \citet{2021AJ....161..100W} suggests that this
approach is not sufficient.
To conclude, both {\it Kepler} and K2 results (after taking the systematic due to light curve length into account) suggest that there is a discrepancy of about 4\%
between the observed and predicted masses, with the observed masses being higher.

\section{Discussion and Conclusions}
In this paper, we have provided an overview and motivation of the K2GAP Guest Observer program, whose main aim is to study the Galaxy through asteroseismology of giants. The program was designed to select stars with an easily reproducible selection function. A total
of 132,197 targets were proposed through K2GAP,
accounting for 30.5\% of the stellar targets observed by K2.
Amongst these, 101,419 stars follow a well defined color-magnitude selection.
One of the most important contributions
of this work is providing rigorous selection function criteria for each campaign in a tabular form. A python code implementing the selection function is also provided.
We also provide a catalog of all stars observed by K2 along with flags to identify stars belonging to our program and stars that strictly satisfy our prescribed selection criteria.
In  order to facilitate comparison with predictions of theoretical Galactic models, we also provide selection-matched mock catalogs generated using {\sl Galaxia}.
We present a simple and efficient ``change point identification'' algorithm used to screen out K2 stars that were proposed by the K2GAP but were serendipitously selected by NASA through other Guest Observer programs.

Our work provides useful guidelines for
designing future astronomy surveys.
Quite often we know the targets that we want
based on some property of the targets.
However, we lack decisive data that measures that property. In such cases it is tempting to design overly sophisticated approaches that in most cases lead to marginal increase in efficiency of selecting the right targets.
We show that in such situations, if possible, adopting a holistic approach can greatly simplify the selection function. For K2GAP, we applied a simple color criteria to focus on the giants. This inevitably led to dwarfs in our sample, which are not useful for doing asteroseismology with K2's 30-min cadence. However, recognizing the fact that they are useful for exoplanet studies, no effort was made to exclude them. This greatly simplified our selection function.

We show that asteroseismic giants in K2 span a wide range of $R$ and $|z|$ in the Galaxy, offering a significant advantage for Galactic studies
compared to \kepler{}.
K2 also contains significantly older stars than \kepler{}, which are useful to probe the early history of the Galaxy.
However, the wider coverage comes at the price of light curves having shorter duration, and consequently lower S/N.

Making use of {\it Gaia} parallaxes, we identify giants that are bright enough to show oscillations
with appreciable SNR and use them to study
$\nu_{\rm max}$ measurement completeness.
We find that for about 14\%
of these giants $\nu_{\rm max}$ cannot be measured.
The exact cause for this is not known and requires further investigation.
Unlike {\it Kepler}, not all stars with
$\nu_{\rm max}$ measurements have a $\Delta \nu$
measurement. The probability to detect
$\Delta \nu$ is maximum at around $\nu_{\rm max}=60$
$\mu$Hz and falls off for higher and lower values of $\nu_{\rm max}$.  This is due to lower SNR and frequency resolution of
K2 compared to \kepler, which makes
detection of $\Delta \nu$ harder. Additionally, our results suggest that  it is
more difficult to detect $\Delta \nu$ for a RC star
compared to an RGB star, most likely due to the oscillation power spectra of RC stars being more complex.

For each campaign, we compare the observed distribution of various  asteroseismic parameters with the predictions of a  Galactic model using {\sl Galaxia}.
Such a comparison is useful to test the model and
theory, but is also useful to check the
quality of the data. This is especially important
for K2 because it collected data over various campaigns with each of them having their unique technical difficulties and challenges over the full course of the mission.
For most campaigns, the observed number of
giants with $\nu_{\rm max}$ measurements is roughly in agreement with predictions, with a observed-to-predicted ratio of stars of 0.78. However, for C11 the ratio is 0.3, which is exceptionally low. Further investigations suggest that this could be due to the roll angle error and the shortened light curve for this particular field.

The distribution of stars in the $(\nu_{\rm max},\kappa_M)$ plane shows a good match with model predictions, both for RGB and RC stars. However,
some differences could also be seen. The
RC sequence in K2 is not as sharp as in {\it Kepler},  due to K2 having higher uncertainty in  $\nu_{\rm max}$ and $\kappa_M$ compared to {\it Kepler}. The location of the RGB-bump was at a lower $\nu_{\rm max}$ in the models compared to the observations as expected from shortcomings in stellar evolution models \citep{2020A&A...635A.164S}. Some RGB stars were found to
be classified as RC in K2. There also seems to
be a lack of secondary RC stars in K2 observations
compared to both the model predictions and the {\it Kepler} results.

We compare the observed $\kappa_M$ distribution with those
of model predictions. We find that for low luminosity giants in K2,
the observed
median $\kappa_M$ is 2.2\% higher than predicted. The observed RC and high luminosity giant distributions differ significantly from predictions. This
is most likely due to significant incompleteness in $\Delta\nu$ measurements.
For low luminosity {\it Kepler} giants the observed
median $\kappa_M$ is 4.1\% higher than predicted. Hence, in general the K2  masses are
lower by about 1.9\% compared to {\it Kepler}, which is in
agreement with findings of \citet{2021MNRAS.506.1761S}
based on stellar kinematics.
As discussed in \citet[see their Figure 5]{2021arXiv210805455Z}, this systematic
offset is due to the shorter time baseline of K2 compared to \kepler.
Hence, data from both K2 and {\it Kepler} suggest
that asteroseismic masses are higher by about 4\% compared to model predictions.
Some of this discrepancy could be due to inaccurate
modelling of $[\alpha/{\rm Fe}]$ enhanced stars
\citep{2021AJ....161..100W}, and some of it could be
due inaccuracies in modelling the Galactic disc
\citet{2021MNRAS.tmp.1920S, 2021MNRAS.506.1761S}.
In future, a further improvement in both the stellar models
and Galactic models is required.

\section{Data availability}
The datasets used are available for download at  \url{http://www.physics.usyd.edu.au/k2gap/}.
The python code for selection function is available at
\url{https://github.com/sanjibs/k2gap}

\section*{Acknowledgements}
We would like to thank the entire community supporting the K2 GAP.
SS is funded by a Senior Fellowship (University of Sydney), an ARC Centre of Excellence for All Sky Astrophysics in 3 Dimensions (ASTRO-3D) Research Fellowship and JBH's Laureate Fellowship from the Australian Research Council (ARC). JBH is supported by an ARC Australian Laureate Fellowship (FL140100278) and ASTRO-3D.
Funding for the Stellar Astrophysics Centre is provided by The Danish National Research Foundation (Grant agreement No.~DNRF106).
Parts of this research were conducted by the Australian Research Council Centre of Excellence for All Sky Astrophysics in 3 Dimensions (ASTRO 3D), through project number CE170100013.

This publication makes use of data products from the Two Micron All Sky Survey (2MASS), which is a joint project of the University of Massachusetts and the Infrared Processing and Analysis Center/California Institute of Technology, funded by the National Aeronautics and Space Administration (NASA) and the National Science Foundation (NSF).

This paper includes data collected by the \kepler{} mission and the K2 mission. Funding for the \kepler{} mission and K2 mission are provided by the NASA Science Mission directorate.

This work has made use of data from the European Space Agency (ESA) mission {\it Gaia} (\url{https://www.cosmos.esa.int/gaia}), processed by the {\it Gaia} Data Processing and Analysis Consortium (DPAC, \url{https://www.cosmos.esa.int/web/gaia/dpac/consortium}). Funding for the DPAC has been provided by national institutions, in particular the institutions participating in the {\it Gaia} Multilateral Agreement.

This research has made use of the VizieR catalogue access tool, CDS,
Strasbourg, France (DOI : 10.26093/cds/vizier). The original description of the VizieR service was published in 2000, A\&AS 143, 23
This research made use of the following software: \texttt{Python~3}, \texttt{numpy} \citep{harris2020array}, \texttt{matplotlib} \citep{hunter2007matplotlib},

\end{document}